\documentclass[10pt,onecolumn]{IEEEtran}
\usepackage{indentfirst}
\usepackage[dvips]{graphicx}
\usepackage{amsfonts}
\usepackage{multirow}
\usepackage{amsmath,amsthm,amssymb}
\usepackage{color,changepage}

\usepackage{CJK}
\usepackage{algorithm}
\usepackage{algorithmic}
\usepackage{bbm}
\usepackage{verbatim}
\usepackage{makecell}
\usepackage{subfigure}
\usepackage{mathrsfs}
\usepackage{arydshln}
\usepackage{cases}
\usepackage{extarrows}
\usepackage{array}

\allowdisplaybreaks[4]

\definecolor{newcolor}{rgb}{0.5,0,1}

\newcommand{\MM}[1]{M}

\newtheorem{Theorem}{Theorem}
\newtheorem{Corollary}{Corollary}
\newtheorem{Definition}{Definition}

\newtheorem{Lemma}{Lemma}

\theoremstyle{remark}
\newtheorem{Remark}{Remark}
\newtheorem{Example}{Example}

\DeclareMathAlphabet{\mathpzc}{OT1}{pzc}{m}{it}

\definecolor{newcolor}{rgb}{0.5,0,1}

\begin{document}

\title{Capacity-Achieving Private Information Retrieval  Schemes from Uncoded Storage Constrained Servers with Low Sub-packetization}
\author{Jinbao Zhu, Qifa Yan, Xiaohu Tang, and Ying Miao
\thanks{J. Zhu, Q. Yan and X. Tang are with the Information Security and National Computing Grid Laboratory, Southwest Jiaotong University, Chengdu 611756, China (email: jinbaozhu@my.swjtu.edu.cn, qifay2014@163.com, xhutang@swjtu.edu.cn).


Y. Miao is with the Faculty of Engineering, Information and Systems, University of Tsukuba, Tennodai 1-1-1, Tsukuba 305-8573, Japan (e-mail: miao@sk.tsukuba.ac.jp).}
}
\maketitle

\begin{abstract}
This paper investigates reducing sub-packetization of capacity-achieving schemes for uncoded Storage Constrained Private Information Retrieval (SC-PIR) systems.
In the SC-PIR system, a user aims to download one out of $K$ files from $N$ servers while revealing nothing about the identity of the requested file to any individual server,
in which the $K$ files are stored at the $N$ servers in an uncoded form and each server can store up to $\mu K$ equivalent files, where $\mu$ is the normalized storage capacity of each server.
 We first prove that there exists a capacity-achieving SC-PIR scheme for a given storage design if and only if all the packets are stored exactly at $M\triangleq \mu N$ servers for $\mu$ such that $M=\mu N\in\{2,3,\ldots,N\}$. Then,  the optimal sub-packetization for capacity-achieving linear SC-PIR schemes is characterized as the solution to an optimization problem, which is typically hard to solve  since it involves non-continuous indicator functions. Moreover,
a new notion of array called \emph{Storage Design Array (SDA)} is introduced  for the SC-PIR system. With any given SDA, an associated  capacity-achieving SC-PIR scheme is constructed. Next, the SC-PIR schemes that have equal-size packets are investigated. Furthermore, the optimal equal-size sub-packetization among all capacity-achieving linear SC-PIR schemes characterized by Woolsey \emph{et al.} is proved to be $\frac{N(M-1)}{\gcd(N,M)}$, which is achieved by a construction of SDA.
Finally, by allowing unequal size of packets,  a greedy SDA construction is proposed, where the sub-packetization of the associated SC-PIR scheme is upper bounded by $\frac{N(M-1)}{\gcd(N,M)}$.
Among all capacity-achieving linear SC-PIR schemes, the sub-packetization is  optimal when $\min\{M,N-M\}|N$ or $M=N$, and within a multiplicative gap $\frac{\min\{M,N-M\}}{\gcd(N,M)}$ of  the optimal one in general. In particular, for  the special case $N=d\cdot M\pm1$ where the positive integer $d\geq 2$, we propose another SDA construction to obtain lower sub-packetization.
\end{abstract}
\begin{IEEEkeywords}
Private information retrieval, uncode, storage constrained servers, sub-packetization, capacity-achieving, storage design array.
\end{IEEEkeywords}

\section{Introduction}\label{Introduction}
Along with the rapid advancement of Distributed Storage Systems (DSSs), protecting the download privacy of a user against public servers is of vital importance. The problem of Private Information Retrieval (PIR) was first introduced  by Chor \emph{et al.} in \cite{Chor} and has attracted remarkable attention within computer science community subsequently \cite{Chor,Gasarch,Ostrovsky,Yekhanin}. In the classical framework, a user wishes to retrieve one out of $K$ files from $N$ servers, each of which stores the whole library of $K$ files, while ensuring that any server can not learn any information about the file index being requested. To this end, the user sends a query string to each server. 
Then the server responds truthfully with an answer string depending on the received query and the contents stored. Finally, the user correctly decodes the requested file from the answers. Note that to prevent each server from obtaining information about which file is being requested, the query distribution has to be marginally independent of the desired file index.

A trivial strategy is to download all the $K$ files in the library no matter which file is requested by the user, but this results in impractical communication cost, especially in a modern DSS, which typically maintains a large number of files.
In the seminal work \cite{Chor} where each file is  of  one bit  size, the communication cost was measured by the sum of upload cost (the total size of query strings) and download cost (the total size of answer strings).
In the sense of information-theoretic security, which assures privacy even if the servers have unbounded computational power,
it was shown in  \cite{Chor} that the naive strategy is the only feasible solution to a single server, whereas low communication cost can be attained by replicating the files at multiple non-colluding servers.
To improve the efficiency, single-server PIR has been widely studied in the sense of computational security, whose privacy is guaranteed by some computational hard problems, for examples, the problems related to so-called $\Phi$-hiding number-theory \cite{Cachin,S-PIR_Gentry}, trapdoor permutations \cite{S-PIR_OT4,trapdoor3}, or quadratic/composite residuosity \cite{N_PIR3,Chang}.
These works improve the efficiency at the cost of non-zero possibility of disclosing information relevant to the identity of the requested file.

Instead of retrieving a single bit,  Shannon theory allows the file size  to be arbitrarily large, and therefore the upload cost can be neglected compared to the download cost since it does not scale with  file size \cite{N. B. Shah,code PIR,Sun replicated,Ulukus_MDS,MDS_Tajeddine2}. 
Then, the communication efficiency  is usually measured by \emph{retrieval rate}, defined as the number of bits that the user can  privately retrieve per bit of download data across all random realizations of queries.  Particularly, the supremum of retrieval rates over all achievable schemes is called \emph{capacity}.
To implement a PIR scheme, the files typically need to be partitioned into  some non-overlapping packets. 
The number of packets is referred to as \emph{sub-packetization} in the literature.
The sub-packetization reflects the complexity of the scheme in practice and is preferred to be as small as possible. This is because any practical scheme will require each of the packets to include some header information  for user to decode \cite{Li Tang}, especially the header overhead may be non-negligible when there are a large number of packets.
This problem has already been noticed in other applications, for example coded caching \cite{Yan:A,Yan:B,Yan,Li Tang,Shan}.

In 2014, Shah \emph{et al.} revisited the PIR problem and reported  an interesting scheme  achieving the PIR rate $1-\frac{1}{N}$ and requiring sub-packetization $N-1$ \cite{N. B. Shah}. Later in  the influential work by Sun and Jafar \cite{Sun replicated}, the exact PIR capacity was characterized as $\big(1+\frac{1}{N}+\ldots+\frac{1}{N^{K-1}}\big)^{-1}$ for any $N$ and $K$.
However, to achieve the capacity, the smallest sub-packetization of the proposed PIR schemes \cite{Sun replicated} is $N^{K}$, which increases exponentially with the number of files $K$ and thus is impractical even for moderate number of files.
Soon afterwards,  the sub-packetization was decreased to $N^{K-1}$ in \cite{Sun optimal}, which was proved to be optimal  under the assumption that the download cost are identical over all random realizations of queries.
In a very recent work  \cite{Tian and Sun}, Tian \emph{et al.} innovatively introduced a new capacity-achieving scheme, which incurs different download cost for distinct realizations of queries.  As a result, the  sub-packetization was decreased to $N-1$,  which is independent of $K$ and  shown to be optimal among all the capacity-achieving PIR schemes.

A common assumption in the aforementioned results is that each server has sufficiently large storage capacity to store all the files in the library, i.e., a repetition coding is used to store the files across servers. Though repetition coding can offer simplicity in designing PIR schemes and the high immunity against server failures, it suffers from  extremely large storage cost. The storage cost in a PIR system has been widely investigated in terms of the coding structures in the storage design, such as specific Maximum Distance Separable (MDS) codes \cite{Ulukus_MDS,MDS_Tajeddine2,Zhu}, an uncoded storage \cite{Tandon Coded caching,Attia SC-PIR,Mingyue Ji}, and other more complicated coding techniques \cite{N. B. Shah,Low storage,Low storage2,Low storage3,Ge_Array code,MDS_Kumar,Banawan ITW}. Moreover, the tradeoff between the storage cost and retrieval rate was considered without any explicit constraints on the storage codes \cite{Sun_storage cost,Tian_storage cost,Tian_storage cost2}. 

As the first step toward exactly characterizing the tradeoff between storage cost and retrieval rate,
Tandon \emph{et al.} formulated the problem of  uncoded Storage Constrained PIR (SC-PIR) in \cite{Tandon Coded caching,Attia SC-PIR}.
In this setup,  each server can store up to $\mu KL$ symbols by some storage design, where $\frac{1}{N}\le\mu\le 1$ is the normalized storage and $L$ is the number of symbols of each file.
The capacity of SC-PIR was proved in \cite{Attia SC-PIR} to be $\big(1+\frac{1}{M}+\ldots+\frac{1}{M^{K-1}}\big)^{-1}$ for $M\triangleq\mu N\in\{1,\ldots,N\}$. However, the capacity-achieving SC-PIR scheme in \cite{Tandon Coded caching,Attia SC-PIR} has sub-packetization $\tbinom{N}{M}M^{K}$. Thus the problem of high sub-packetization  shows up again in this SC-PIR model.
Recently, Woolsey \emph{et al.} \cite{Mingyue Ji} proposed a general construction of SC-PIR schemes by establishing the connection between storage design and Storage Full PIR (SF-PIR, i.e., the case that each server can store all the files).   Then, the sub-packetization to achieve the capacity of the SC-PIR system was reduced to $NM^{K-1}$ in \cite{Mingyue Ji}, which also increases exponentially with $K$. 

In this paper, we are interested in characterizing the optimal sub-packetization to achieve the capacity of SC-PIR systems.
Note from the previous work \cite{Attia SC-PIR,Mingyue Ji} that \emph{linear} schemes are sufficient to achieve the capacity of SC-PIR.
Additionally, it was proved in \cite{Attia SC-PIR} that, for any $\mu$ with $\frac{1}{N}\leq \mu\leq 1$, the capacity of SC-PIR system can be achieved by memory-sharing technique between the discrete points such that $M\in\{1,2,\ldots,N\}$, where $M=1$ is a trivial case since the user has to download all the contents stored at the $N$ servers to assure privacy. Therefore, the problem comes down to the case $M\in\{2,3,\ldots,N\}$ for  linear SC-PIR schemes, which is the focus of this paper. The contributions of this paper are:
\begin{enumerate}
\item 
      We prove that there exists a capacity-achieving SC-PIR scheme for a given storage design if and only if all the packets are stored exactly at $M$ servers in the storage phase.
  \item We characterize the optimal sub-packetization of capacity-achieving linear SC-PIR schemes by an optimization problem.
   Consequently, a general construction of capacity-achieving linear SC-PIR schemes with optimal sub-packetization can be obtained based on  the optimal solution of this optimization problem.
  \item Storage Design Array (SDA) is introduced to obtain feasible solutions to the optimization problem. Any given SDA is associated to a practical capacity-achieving linear SC-PIR scheme with low sub-packetization.
  \item 
We prove that the optimal \emph{equal-size} sub-packetization is $\frac{N(M-1)}{\gcd(N,M)}$ among all the classes of capacity-achieving linear SC-PIR schemes characterized by Woolsey \emph{et al.} \cite{Mingyue Ji}.
  \item In order to further decrease sub-packetization of capacity-achieving SC-PIR schemes, we investigate the problem under a more general assumption, i.e., the sizes of the packets are allowed to be unequal.
      In particular, a greedy algorithm is proposed to construct SDA for any positive integers $N,M$ such that $1\leq M\leq N$.
      The sub-packetization of the associated SC-PIR scheme is shown to be optimal among all capacity-achieving linear SC-PIR schemes when $\min\{M,N-M\}|N$ or $M=N$. In the other cases, the sub-packetization is within a multiplicative gap $\frac{\min\{M,N-M\}}{\gcd(N,M)}$ compared to its lower bound.
      Moreover, for the case $N=d\cdot M\pm1$ where the integer $d\geq 2$, we propose another construction of SDA to achieve lower sub-packetization compared to the greedy SDA.
 \end{enumerate}

The rest of this paper is organized as follows. In Section \ref{system model}, we introduce the system model and problem formulation.
In Section \ref{lower bound:sub-pack}, we establish an information-theoretic lower bound on sub-packetization of capacity-achieving linear SC-PIR schemes.
In Section \ref{packets:charact}, we characterize a generic construction of capacity-achieving linear SC-PIR schemes with optimal sub-packetization.
In Section \ref{SDA:def}, we introduce SDA to construct capacity-achieving SC-PIR schemes with low sub-packetization.
In Section \ref{equal-size packets}, we present the results under the assumption of equal-size packets.
Section \ref{schemes SDA} proposes two SDA constructions and proves the optimality of the resultant sub-packetization.
Finally, the paper is concluded in Section \ref{conclusion}.

The following notation is used throughout this paper.
\begin{itemize}
  \item For any integers $n,m,s,N$ with $n\leq m$, $[n:m]$ and $([n:m]+s)_N$ respectively denote the sets $\{n,n+1,\ldots,m\}$ and $\{i+s \,(\bmod\, N):n\le i\le m\}$;
  \item For a finite set $\mathcal{S}$, $|\mathcal{S}|$ denotes its cardinality;
 \item Denote $A_{1:m}$ a vector $(A_{1},\ldots,A_{m})$, and define $A_{\Gamma}$ as $(A_{\gamma_1},\ldots,A_{\gamma_{k}})$ for any index set $\Gamma=\{\gamma_1,\ldots,\gamma_{k}\}\subseteq[1:m]$ with $\gamma_1<\ldots<\gamma_k$ or any index vector $\Gamma=(\gamma_1,\ldots,\gamma_{k})$;
 \item Define $\mathbf{1}(x)$ as a function of a logical variable $x$, i.e., $\mathbf{1}(x)=1$ if  $x$ is true and $\mathbf{1}(x)=0$ otherwise.
  \end{itemize}

\section{System Model}\label{system model}
Let  $\mathbb{F}_{q}$ be  the finite field for a prime power $q$.
Consider a non-colluding PIR system with $K$ files  $W_1,\ldots,W_{K}\in\mathbb{F}_{q}^{L\times 1}$  stored across $N$ 
servers in an uncoded fashion. Each of files is comprised of $L$ i.i.d. uniform symbols over $\mathbb{F}_{q}$, i.e.,
\begin{IEEEeqnarray}{rCl}
H( W_1)&=&\ldots=H( W_{K})=L, \label{infor indenpe}\\
H( W_1,\ldots,W_{K})&=&\sum_{k=1}^{K}H(W_k), \label{model:file inden}
\end{IEEEeqnarray}
where the entropy function $H(\cdot)$ is measured with logarithm $q$.
Let $Z_n$ ($n\in[1:N]$) be the contents stored at server $n$, which is subject to the storage capacity of server $n$, then the \emph{storage constraint} for each server is
\begin{IEEEeqnarray}{c}
H(Z_{n})\leq\mu KL,\quad\forall \,n\in[1:N],\label{storage constraint}
\end{IEEEeqnarray}
where $\mu$ is the \emph{normalized storage capacity}. Notice that, when $\mu<\frac{1}{N}$, the total storage capacity of $N$ servers is insufficient to store all the $K$ files. For $\mu=1$, each server can store all the $K$ files. Thus, we are interest in the case $\frac{1}{N}\leq \mu\leq 1$.

The system operates in the following two phases:

\textbf{Storage Phase:}
Each file $W_k$ is partitioned into $F$ disjoint \emph{packets} and thus it will be convenient to label
the $F$ packets as $W_{k,1},W_{k,2},\ldots,W_{k,F}$, where $W_{k,i}$ is the $i$-th packet of file $W_k$. By convention, we call $F$ \emph{sub-packetization}. Then, for any  $k\in[1:K]$,
\begin{IEEEeqnarray}{rCl}
W_k&=&\left\{W_{k,i}:i\in[1:F]\right\},\label{file:partition}\\
H(W_k)&=&\sum_{i=1}^FH(W_{k,i}).\label{packets:independent}
\end{IEEEeqnarray}

Clearly, each of these packets must be stored at \emph{at least one server} because of the constraint of reliable decoding. In particular, all the files are partitioned and stored in the same manner\footnote{To the best of our knowledge, all the previous storage constrained PIR schemes satisfy this assumption \cite{Tandon Coded caching,Attia SC-PIR,Mingyue Ji}, which is also a popular storage manner in coded caching \cite{Maddah-Ali,Yan,Li Tang,Shan,Maddah-Ali2}.}, i.e.,
\begin{IEEEeqnarray}{rl}
H(W_{1,i})=H(W_{2,i})=\ldots=H(W_{K,i}),&\quad\forall \,i\in[1:F],\label{file:same_entropy}\\
Z_n=\{W_{k,i}:k\in[1:K],i\in \mathcal{Z}_n\},&\quad \forall\, n\in[1:N],\label{file:same_storage}
\end{IEEEeqnarray}
where $\mathcal{Z}_n$ is a subset of $[1:F]$ such that $Z_n$ satisfies \eqref{storage constraint}. In other words, $\mathcal{Z}_n\subseteq[1:F]$ consists of the indices of packets stored at server $n$.

\textbf{Retrieval Phase:} A user selects an index $\theta\in[1:K]$  privately and wishes to retrieve the file $W_{\theta}$ from the system without disclosing any information about $\theta$ to  any individual server. For this purpose, the user generates $N$ queries $Q_{1:N}^{[\theta]}$ and sends $Q_{n}^{[\theta]}$ to server $n\in [1:N]$. Indeed, the queries are generated independently of file realizations, i.e.,
\begin{IEEEeqnarray}{c}\label{model:query inden}
I({Q}_{1:N}^{[\theta]};{W}_{1:K})=0,\quad\forall\, \theta\in[1:K],
\end{IEEEeqnarray}
where $I(\cdot)$ is the mutual information function.
Upon receiving the query $Q_{n}^{[\theta]}$, server $n$ responds with an answer $A_{n}^{[\theta]}$, which is determined by the received query and its stored contents. Thus, by the data processing inequality,
\begin{IEEEeqnarray}{c}\label{model:answers}
H(A_{n}^{[\theta]}|Q_{n}^{[\theta]},Z_{n})=H(A_{n}^{[\theta]}|Q_{n}^{[\theta]},{W}_{1:K})=0,\quad\forall\, n\in[1:N].
\end{IEEEeqnarray}
Finally, from all the answers $A_{1:N}^{[\theta]}$ collected from the $N$ servers, the user must be able to decode the desired file $W_{\theta}$ correctly, i.e.,
\begin{IEEEeqnarray}{c}\label{model:decod const}
H(W_{\theta}|A_{1:N}^{[\theta]},Q_{1:N}^{[\theta]})=0,\quad\forall\, \theta\in[1:K].
\end{IEEEeqnarray}
To ensure the privacy, the strategies for retrieving any two files $W_{\theta}$ and $W_{\theta'}$ must be indistinguishable in terms of any individual server, i.e.,
\begin{IEEEeqnarray}{c}\label{model:ident dis}
(Q_{n}^{[\theta]},A_{n}^{[\theta]},Z_{n})\sim (Q_{n}^{[\theta']},A_{n}^{[\theta']},Z_{n}),\quad\forall\, \theta,\theta'\in[1:K], \forall\, n\in[1:N],
\end{IEEEeqnarray}
where $X\sim Y$ means that the random variables $X$ and $Y$ are  identical distribution. Equivalently, the desired index $\theta$ must be hidden from all the information available to each server, i.e.,
\begin{IEEEeqnarray}{c}\label{Infor:priva cons}
I(Q_{n}^{[\theta]},A_{n}^{[\theta]},Z_{n};\theta)=0,\quad\forall\, n\in[1:N].
\end{IEEEeqnarray}

Throughout this paper, we refer to this system as a  $(\mu,N,K)$ \emph{Storage Constrained 
PIR} (SC-PIR) system. If $\mu=1$,  the system is also referred to as an $(N,K)$ \emph{Storage Full PIR} (SF-PIR) system.

In order to measure the performance of SC-PIR systems, the following two quantities are considered:
\begin{enumerate}
  \item[1.] The sub-packetization $F$, which reflects the complexity of the SC-PIR scheme in practical applications, and thus is preferred to be as small as possible.
  \item[2.] The retrieval rate $R$, which is the number of desired bits that the user can retrieve privately per bit of downloaded data, is defined as
        \begin{IEEEeqnarray}{c}\label{def:rate}
        R\triangleq\frac{H(W_{\theta})}{\sum_{n=1}^{N}H(A_{n}^{[\theta]})}=\frac{L}{D},
        \end{IEEEeqnarray}
        where $D\triangleq \sum_{n=1}^{N}H(A_{n}^{[\theta]})$ is the average download cost from the $N$ servers over random queries. Obviously, $R$ and $D$ are independent of $\theta$ by  \eqref{infor indenpe} and \eqref{model:ident dis}.

        A retrieval rate $R$ is said to be achievable if there exists a design of both storage and retrieval phases satisfying \eqref{storage constraint}--\eqref{Infor:priva cons}  such that its retrieval rate is greater than or equal to $R$.
        The capacity of the SC-PIR system, denoted by $C^*$, is the supremum over all the achievable rates, i.e.,
        \begin{IEEEeqnarray}{c}\notag
        C^*=\sup\left\{R: R ~\text{is achievable} \right\}.
        \end{IEEEeqnarray}
\end{enumerate}

Define the \emph{total normalized storage capacity} as
 \begin{IEEEeqnarray}{c}\notag
 M\triangleq\mu N\in[1,N].\label{def:T}
 \end{IEEEeqnarray}
 For the case $M\in[1:N]$,  the capacity of SC-PIR is exactly characterized in \cite{Attia SC-PIR} as
\begin{IEEEeqnarray}{c}\label{optimal download}
C^*=\left( 1+\frac{1}{M}+\ldots+\frac{1}{M^{K-1}} \right)^{-1}.
\end{IEEEeqnarray}

Generally, for other $M\in[1,N]$ (or equivalently $\mu\in[\frac{1}{N},1]$), the capacity  can be achieved by memory-sharing technique between the integer points $\lceil M\rceil$ and $\lfloor M\rfloor$ (see \cite[Claim 1 \& Theorem 2]{Attia SC-PIR}). Thus, in the sequel, we will concentrate our discussion on the case $M\in[2:N]$ since it is straightforward to prove that the optimal sub-packetization is $F=N$ for the case $M=1$.

Moreover, the existing work \cite{Tandon Coded caching,Attia SC-PIR,Mingyue Ji} have shown that \emph{linear SC-PIR schemes} can achieve the capacity.
\begin{Definition}[Linear SC-PIR Scheme]\label{def:linear}
For a given scheme of the $(\mu,N,K)$ SC-PIR system, let $\ell_n$ be the \emph{answer length}\footnote{Throughout this paper, the ``length" is counted by the number of packets, thus ``answer length" refers to as the number of packets  in the answer.} of query $Q_{n}^{[\theta]}$. It is said to be a  \emph{linear SC-PIR scheme} if the answers $A_{n}^{[\theta]}$ $(n\in[1:N])$ are formed by
\begin{IEEEeqnarray}{c}\notag
A_{n}^{[\theta]}=\mathbf{LC}_{n}^{[\theta]} (Z_n)=\left( \mathbf{LC}_{n,1}^{[\theta]} (Z_n),\ldots,\mathbf{LC}_{n,\ell_n}^{[\theta]} (Z_n) \right),\quad\forall\, n\in[1:N]
\end{IEEEeqnarray}
with each entry $\mathbf{LC}_{n,j}^{[\theta]} (Z_n)$ ($j\in[1:\ell_n]$) given by a linear combination of the packets stored at server $n$, i.e.,
\begin{IEEEeqnarray}{c}\label{def:coefficients}
\emph{\textbf{LC}}_{n,j}^{[\theta]} (Z_n)=
\sum\limits_{k\in[1:K]}\sum_{i\in\mathcal{Z}_n}\beta_{n,k,i,j}^{[\theta]}\cdot W_{k,i},
\end{IEEEeqnarray}
where $\beta_{n,k,i,j}^{[\theta]}\in\mathbb{F}_q$ is the coefficient of packet $W_{k,i}$ in the $j$-th entry of $A_n^{[\theta]}$ and is \emph{determined completely} by the received query $Q_{n}^{[\theta]}$. Here, it implicitly assumes that each of packets $\{W_{k,i}:k\in[1:K],i\in[1:F]\}$ is represented by a vector over $\mathbb{F}_q$.
If the packets have different  dimensions, then the additions are performed by padding the vectors with zeros to the largest dimension.
\end{Definition}

The objective of the paper is to design $(\mu,N,K)$ linear SC-PIR schemes achieving the SC-PIR capacity with the minimum sub-packetization for the case $M\in[2:N]$.

\section{A Lower Bound on Sub-packetization of Capacity-Achieving Linear SC-PIR Schemes}\label{lower bound:sub-pack}
To simplify our notations in the following discussion, denote $W_{k,\mathcal{S}}$  the \emph{set of packets} of file $W_k$ that are \emph{exclusively} stored by servers in $\mathcal{S}$,  $\mathcal{S}\subseteq [1:N]$, i.e.,
\begin{IEEEeqnarray}{c}\label{def:W_ks}
W_{k,\mathcal{S}}\triangleq \left\{W_{k,i}:i\in\left(\mathop\cap_{n\in\mathcal{S}}  \mathcal{Z}_n\right)\bigg\backslash\left(\mathop\cup_{m\in [1:N]\backslash\mathcal{S}} \mathcal{Z}_m\right)\right\},\quad\forall\, k\in[1:K].
\end{IEEEeqnarray}
Obviously, $W_{k,\emptyset}=\emptyset$ and $H(W_{k,\emptyset})=0$ due to the constraint of reliable decoding.
Then, file $W_k$  and the storage contents at server $n$ can be respectively  rewritten as
\begin{IEEEeqnarray}{rCl}
W_{k}&=&\mathop\cup\limits_{\mathcal{S}\subseteq[1:N]}W_{k,\mathcal{S}}, \quad\forall\, k\in[1:K]\label{file:express}
\end{IEEEeqnarray}
and
\begin{IEEEeqnarray}{rCl}
Z_n&=&\mathop\cup\limits_{k\in[1:K]}\mathop\cup\limits_{\substack{\mathcal{S}\subseteq[1:N]\\n\in\mathcal{S}}}W_{k,\mathcal{S}},\quad\forall\, n\in[1:N].\label{storage place}
\end{IEEEeqnarray}

Notice from \eqref{file:same_entropy} and \eqref{file:same_storage} that both the entropy of random variable $W_{k,\mathcal{S}}$ and the size of set $W_{k,\mathcal{S}}$  are irrespective of $k$.
Thus, for all $\mathcal{S}\subseteq[1:N]$,
we can set $H(W_{k,\mathcal{S}})\triangleq\alpha_{\mathcal{S}}L$ and $F_{\mathcal{S}}\triangleq|W_{k,\mathcal{S}}|$ where $\alpha_{\mathcal{S}}\in[0,1]$.
In other words, $\alpha_{\mathcal{S}}$ is the normalized file size of $W_{k,\mathcal{S}}$  and $F_{\mathcal{S}}$ is the number of packets in $W_{k,\mathcal{S}}$.
By \eqref{infor indenpe}, \eqref{storage constraint}, \eqref{file:express} and \eqref{storage place},  the file size, storage size, and sub-packetization $F$ are respectively constrained as
\begin{IEEEeqnarray}{rCl}
\sum\limits_{\mathcal{S}\subseteq[1:N]}\alpha_{\mathcal{S}}&=&1,\label{const:file size} \\
\sum\limits_{\substack{\mathcal{S}\subseteq[1:N]\\n\in\mathcal{S}}}\alpha_{\mathcal{S}}&\leq &\mu,\quad\forall\, n\in[1:N],\label{uncode:storage constraint}
\end{IEEEeqnarray}
and
\begin{IEEEeqnarray}{c}\label{def:sub-packetization}
F=\sum_{\mathcal{S}\subseteq[1:N]}F_{\mathcal{S}}.
\end{IEEEeqnarray}

In the following, we establish an information-theoretical lower bound on sub-packetization of any capacity-achieving $(\mu,N,K)$ linear SC-PIR scheme with $M=\mu N\in[2:N]$,
which is characterized by the following optimization problem. 
\begin{Definition} Given any positive integers $N$ and $M=\mu N\in[2:N]$, \textbf{Problem 1} is defined as
\begin{IEEEeqnarray}{ccLl}
 \{\alpha_{\mathcal{S}}^{*}\}_{\mathcal{S}\subseteq[1:N],|\mathcal{S}|=M}=\;& \arg\min &\quad \sum\limits_{\substack{\mathcal{S}\subseteq[1:N]\\|\mathcal{S}|=M}}\mathbf{1}(\alpha_{\mathcal{S}}>0)& \notag \\
& s.t. &\quad \sum\limits_{\substack{\mathcal{S}\subseteq[1:N]\\|\mathcal{S}|=M, n\in\mathcal{S}}}\alpha_{\mathcal{S}}=\mu,&\quad\forall\, n\in[1:N] \label{optimize:condition1} \\
& &\quad 0\leq\alpha_{\mathcal{S}}\leq 1,&\quad\forall\,\mathcal{S}\subseteq[1:N], |\mathcal{S}|=M\label{optimize:condition2}
\end{IEEEeqnarray}
where  $\{\alpha_{\mathcal{S}}^{*}\}_{\mathcal{S}\subseteq[1:N],|\mathcal{S}|=M}$ is called the optimal solution to Problem 1 and $\eta^*=\sum\limits_{\substack{\mathcal{S}\subseteq[1:N],|\mathcal{S}|=M}}\mathbf{1}(\alpha_{\mathcal{S}}^*>0)$ is
called the optimal value of Problem 1. In addition, the parameters $\{\alpha_{\mathcal{S}}\}_{\mathcal{S}\subseteq[1:N],|\mathcal{S}|=M}$  satisfying \eqref{optimize:condition1} and \eqref{optimize:condition2} are called a feasible solution to Problem 1.
\end{Definition}

\subsection{Necessary Conditions of Capacity-Achieving Linear SC-PIR Schemes}
In this subsection, we derive five necessary conditions (Lemmas \ref{necessary:P1-2} and \ref{lemma:necess} below) for capacity-achieving linear SC-PIR schemes, whose proofs  are  left in Appendix. 

\begin{Lemma}\label{necessary:P1-2}
Given any $(\mu,N,K)$ SC-PIR system with $M=\mu N\in[2:N]$ and $\{\alpha_{\mathcal{S}}: \alpha_{\mathcal{S}}\in [0,1], \mathcal{S}\subseteq[1:N]\}$, the storage design of any capacity-achieving SC-PIR scheme must satisfy:
\begin{enumerate}
  \item[P1.] All the packets must be stored exactly at $M$ servers, i.e., $\alpha_{\mathcal{S}}=0$ for all $\mathcal{S}\subseteq[1:N]$ with $|\mathcal{S}|\neq M$;
  \item[P2.] The storage capacity at all servers must be used up, i.e., $\sum\limits_{\mathcal{S}\subseteq[1:N],n\in\mathcal{S}}\alpha_{\mathcal{S}}=\mu$ for all $n\in[1:N]$.
\end{enumerate}
\end{Lemma}
\begin{Remark}\label{remark:equivalent}
Given any parameters $\{\alpha_{\mathcal{S}}: \alpha_{\mathcal{S}}\in [0,1], \mathcal{S}\subseteq[1:N]\}$, P1 along with P2 are equivalent to the constraints \eqref{optimize:condition1} and \eqref{optimize:condition2} of Problem 1.
\end{Remark}

For any $\mathcal{K}\subseteq[1:K],\mathcal{S}\subseteq[1:N]$, denote $W_{\mathcal{K},\mathcal{S}}\triangleq\mathop\cup\limits_{k\in\mathcal{K}}W_{k,\mathcal{S}}$. Given $\theta\in[1:K]$, let $\widetilde{\mathbf{LC}}_{n}^{[\theta]}(Z_n)$ be the answer of server $n$ when receiving the query realization $\widetilde{Q}_{n}^{[\theta]}$.
Let $\widetilde{\mathbf{LC}}_n^{[\theta]}(W_{\mathcal{K},\mathcal{S}})$ be the part of $\widetilde{\mathbf{LC}}_{n}^{[\theta]} (Z_n)$ involving the linear combinations of packets in $W_{\mathcal{K},\mathcal{S}}$, i.e.,
 \begin{IEEEeqnarray}{c}\label{add:notation1}
 \widetilde{\mathbf{LC}}_n^{[\theta]}(W_{\mathcal{K},\mathcal{S}})\triangleq \left( \widetilde{\mathbf{LC}}_{n,1}^{[\theta]} (W_{\mathcal{K},\mathcal{S}}),\ldots,\widetilde{\mathbf{LC}}_{n,\widetilde{\ell}_n}^{[\theta]}(W_{\mathcal{K},\mathcal{S}}) \right),\quad\forall\, n\in[1:N],
\end{IEEEeqnarray}
where $\widetilde{\ell}_n$ is the answer length for the query realization $\widetilde{Q}_{n}^{[\theta]}$, and $\widetilde{\mathbf{LC}}_{n,j}^{[\theta]} (W_{\mathcal{K},\mathcal{S}})$ is given by
 \begin{IEEEeqnarray}{c}\label{add:notation2}
 \widetilde{\mathbf{LC}}_{n,j}^{[\theta]} (W_{\mathcal{K},\mathcal{S}})=
\sum\limits_{k\in \mathcal{K}, W_{k,i}\in W_{\mathcal{K},\mathcal{S}}}\widetilde{\beta}_{n,k,i,j}^{[\theta]}\cdot W_{k,i},\quad\forall\, j\in[1:\widetilde{\ell}_n]
 \end{IEEEeqnarray}
in which the coefficient $\widetilde{\beta}_{n,k,i,j}^{[\theta]}$ is the realization of $\beta_{n,k,i,j}^{[\theta]}$ in \eqref{def:coefficients} when the query realization $\widetilde{Q}_{n}^{[\theta]}$ is received by server $n$. 

\begin{Lemma}\label{lemma:necess}
Given any $(\mu, N,K)$ SC-PIR system  with $M=\mu N\in[2:N]$, let $\mathcal{S}\subseteq[1:N]$ and $\theta,\theta'\in[1:K]$ such that $|\mathcal{S}|=M,\theta\neq\theta'$.
For every  realization of queries $\widetilde{Q}_{1:N}^{[\theta]}$  with positive probability, the retrieval phase for any capacity-achieving linear SC-PIR scheme must satisfy:
\begin{enumerate}
  \item[P3.] \emph{(Independence of the retrieved data)} The $M$ random variables
                \begin{IEEEeqnarray}{c}\label{condition:P3}
                \widetilde{\emph{\textbf{LC}}}_{n}^{[\theta]}\big(W_{\theta,\mathcal{S}}\big),\quad\forall\, n\in\mathcal{S}
                \end{IEEEeqnarray}
                are independent of each other;
  \item[P4.] \emph{(Independence of the requested data)} The $M$ random variables
                \begin{IEEEeqnarray}{c}\label{condition:P5}
                \widetilde{\mathbf{LC}}_{n}^{[\theta]}\big(W_{[1:K]\backslash\{\theta'\},\mathcal{S}}\big),\quad\forall\, n\in\mathcal{S}
                \end{IEEEeqnarray}
                are independent of each other;
  \item[P5.] \emph{(Identical information for the residuals)} The $M$ random variables
                  \begin{IEEEeqnarray}{c}\label{condition:P4}
                  \widetilde{\emph{\textbf{LC}}}_{n}^{[\theta]}\big(W_{[1:K]\backslash\{\theta,\theta'\},\mathcal{S}}\big),\quad\forall\, n\in\mathcal{S}
                  \end{IEEEeqnarray}
                  are deterministic of each other.
\end{enumerate}
\end{Lemma}

\subsection{Lower Bound on Sub-packetization of Capacity-Achieving Linear SC-PIR Schemes}

\begin{Lemma}\label{bound:each subfile}
Given any capacity-achieving $(\mu,N,K)$ linear SC-PIR scheme with $M=\mu N\in[2:N]$ and $\{\alpha_{\mathcal{S}}: \alpha_{\mathcal{S}}\in [0,1], \mathcal{S}\subseteq[1:N]\}$,
\begin{IEEEeqnarray}{c}\notag
\left\{\begin{array}{@{}ll}
F_{\mathcal{S}}\geq M-1,   &\mathrm{if}~\alpha_{\mathcal{S}}>0, \mathcal{S}\subseteq[1:N],|\mathcal{S}|=M  \\
F_{\mathcal{S}}=0,&\mathrm{otherwise}
\end{array}\right..
\end{IEEEeqnarray}
\end{Lemma}

\begin{IEEEproof}
It is clear that $F_{\mathcal{S}}=0$ if $\alpha_{\mathcal{S}}=0$. By Lemma \ref{necessary:P1-2}, we just need to prove $F_{\mathcal{S}}\geq M-1$ if $\alpha_\mathcal{S}>0,\mathcal{S}\subseteq[1:N],|\mathcal{S}|=M$.

Let $W_{\theta^{*}}$ and $\widetilde{Q}_{1:N}^{[\theta^{*}]}$ be the desired file of the user and a realization of queries with positive probability, respectively.
For any $\mathcal{S}$ with $|W_{\theta^{*},\mathcal{S}}|>0$, recall from \eqref{def:W_ks} that $W_{\theta^{*},\mathcal{S}}$  are exclusively stored at servers in $\mathcal{S}$. Thus, in the conditioning of the realization of queries $\widetilde{Q}_{1:N}^{[\theta^{*}]}$, to ensure that the user can correctly decode $W_{\theta^{*}}$, there must be a server $n\in\mathcal{S}$ such that the coefficients of packets $W_{\theta^{*},\mathcal{S}}$ in $\widetilde{\textbf{LC}}_{n}^{[\theta^{*}]}(Z_{n})$ are not all zeros, i.e.,
\begin{IEEEeqnarray}{c}\notag
H\Big(\widetilde{\textbf{LC}}_{n}^{[\theta^{*}]}(W_{\theta^{*},\mathcal{S}})\Big)>0,\quad\forall\, n\in\mathcal{S}.
\end{IEEEeqnarray}

Note  that, the random queries for retrieving distinct files at a given server have the identical distribution  by  the privacy constraint \eqref{model:ident dis}. Thus,  the following observation holds:
\begin{adjustwidth}{20pt}{0cm}
\textbf{Observation: }For any realization of queries $\widetilde{Q}_{1:N}^{[\theta^{*}]}$ with positive probability, the query $\widetilde{Q}_{n}^{[\theta^{*}]}$, sent to server $n$ for retrieving file $W_{\theta^{*}}$, can also be sent to the same server $n$ but for retrieving any distinct file $\theta\neq\theta^{*}$ in another realization of queries $\widetilde{{Q}}_{1:N}^{[\theta]}$ with positive probability, where $\widetilde{Q}_{n}^{[\theta]}=\widetilde{{Q}}_{n}^{[\theta^{*}]}$. As a result, for the two realizations of queries $\widetilde{Q}_{1:N}^{[\theta^{*}]}$ and $\widetilde{Q}_{1:N}^{[\theta]}$, server $n$ will respond  the same answer, i.e.,
\begin{IEEEeqnarray}{c}\notag
\widetilde{\textbf{LC}}_{n}^{[\theta^{*}]}(Z_n)=\left( \widetilde{\textbf{LC}}_{n,1}^{[\theta^{*}]} (Z_n),\ldots,\widetilde{\textbf{LC}}_{n,\widetilde{\ell}_n}^{[\theta^{*}]} (Z_n) \right)=
\left( \widetilde{\textbf{LC}}_{n,1}^{[\theta]} (Z_n),\ldots,\widetilde{\textbf{LC}}_{n,\widetilde{\ell}_n}^{[\theta]} (Z_n) \right)=\widetilde{\textbf{LC}}_{n}^{[\theta]}(Z_n).
\end{IEEEeqnarray}
\end{adjustwidth}

That is, for any $\theta\in[1:K]\backslash\{\theta^{*}\}$, there exists another realization of queries $\widetilde{Q}_{1:N}^{[\theta]}$ with positive probability such that server $n$ will respond with the same answer
$\widetilde{\textbf{LC}}_{n}^{[\theta]}(Z_{n})=\widetilde{\textbf{LC}}_{n}^{[\theta^{*}]}(Z_{n})$, where the terms involving $W_{\theta^{*},\mathcal{S}}$ are identical, i.e.,
\begin{IEEEeqnarray}{c}\notag
H\Big(\widetilde{\textbf{LC}}_{n}^{[\theta]}(W_{\theta^{*},\mathcal{S}})\Big)=H\Big(\widetilde{\textbf{LC}}_{n}^{[\theta^{*}]}(W_{\theta^{*},\mathcal{S}})\Big)>0.
\end{IEEEeqnarray}
Then, for any $\theta'\in[1:K]\backslash\{\theta,\theta^{*}\}$,
\begin{IEEEeqnarray}{rCl}
&&H\Big(\widetilde{\textbf{LC}}_{n}^{[\theta]}(W_{[1:K]\backslash\{\theta,\theta'\},\mathcal{S}})\Big) \notag\\
&\overset{(a)}{\geq}&H\Big(\widetilde{\textbf{LC}}_{n}^{[\theta]}(W_{[1:K]\backslash\{\theta,\theta'\},\mathcal{S}})\big|W_{[1:K]\backslash\{\theta,\theta',\theta^{*}\},\mathcal{S}}\Big) \notag \\
&\overset{(b)}{=}&H\Big(\widetilde{\textbf{LC}}_{n}^{[\theta]}(W_{\theta^{*},\mathcal{S}})\Big) \notag\\
&>& 0, \label{anser:non zero}
\end{IEEEeqnarray}
where $(a)$ holds because conditioning reduces entropy; $(b)$ follows from the linearity of \eqref{add:notation2}.

Assume that the number of packets in $W_{\theta,\mathcal{S}}$ is less than $M-1$, i.e.,  $F_{\mathcal{S}}<M-1$. According to  \eqref{add:notation2} and \eqref{condition:P3}, the $M$ random variables $\widetilde{{\textbf{LC}}}_{n'}^{[\theta]}\big(W_{\theta,\mathcal{S}}\big)$  ($n'\in\mathcal{S}$) consisting of linear combinations of $F_{\mathcal{S}}$ packets in $W_{\theta,\mathcal{S}}$ are independent of each other. Thus, $F_{\mathcal{S}}<M-1$ results in that there must exist two distinct servers $i,j\in \mathcal{S}$ such that
\begin{IEEEeqnarray}{c}\label{desired:zero}
\widetilde{{\textbf{LC}}}_{i}^{[\theta]}(W_{\theta,\mathcal{S}})=\widetilde{{\textbf{LC}}}_{j}^{[\theta]}(W_{\theta,\mathcal{S}})=\mathbf{0},\quad\forall\, i,j\in\mathcal{S},i\neq j.
\end{IEEEeqnarray}
However,  we have
\begin{IEEEeqnarray}{rCl}
0&\overset{(a)}{=}&I\Big(\widetilde{\textbf{LC}}_{i}^{[\theta]}(W_{[1:K]\backslash\{\theta'\},\mathcal{S}});\widetilde{\textbf{LC}}_{j}^{[\theta]}(W_{[1:K]\backslash\{\theta'\},\mathcal{S}})\Big) \notag\\
&\overset{(b)}{=}&I\Big(\widetilde{\textbf{LC}}_{i}^{[\theta]}(W_{[1:K]\backslash\{\theta',\theta\},\mathcal{S}})+\widetilde{\textbf{LC}}_{i}^{[\theta]}(W_{\theta,\mathcal{S}});
\widetilde{\textbf{LC}}_{j}^{[\theta]}(W_{[1:K]\backslash\{\theta',\theta\},\mathcal{S}})+\widetilde{\textbf{LC}}_{j}^{[\theta]}(W_{\theta,\mathcal{S}})\Big) \notag \\
&\overset{(c)}{=}&I\Big(\widetilde{\textbf{LC}}_{i}^{[\theta]}(W_{[1:K]\backslash\{\theta',\theta\},\mathcal{S}});\widetilde{\textbf{LC}}_{j}^{[\theta]}(W_{[1:K]\backslash\{\theta',\theta\},\mathcal{S}})\Big) \notag \\
&\overset{(d)}{=}&H\Big(\widetilde{\textbf{LC}}_{i}^{[\theta]}(W_{[1:K]\backslash\{\theta',\theta\},\mathcal{S}})\Big) \notag \\
&\overset{(e)}{>}&0, \notag
\end{IEEEeqnarray}
where $(a)$ follows by \eqref{condition:P5}; $(b)$ follows from the linearity of \eqref{add:notation2} again; $(c)$ is due to \eqref{desired:zero}; $(d)$ is because of \eqref{condition:P4}; $(e)$ holds since
\begin{IEEEeqnarray}{c}\notag
H\Big(\widetilde{\textbf{LC}}_{i}^{[\theta]}(W_{[1:K]\backslash\{\theta,\theta'\},\mathcal{S}})\Big)=H\Big(\widetilde{\textbf{LC}}_{n}^{[\theta]}(W_{[1:K]\backslash\{\theta,\theta'\},\mathcal{S}})\Big)
>0
\end{IEEEeqnarray}
by \eqref{condition:P4} and \eqref{anser:non zero}. Thus, $F_{\mathcal{S}}\geq M-1$ and the proof is completed.
\end{IEEEproof}

Now, we are ready to characterize a lower bound on the  sub-packetization among all capacity-achieving linear SC-PIR schemes.
\begin{Theorem}\label{arbitra:part}
For any given $(\mu,N,K)$ SC-PIR system with $M=\mu N\in[2:N]$, the sub-packetization of any capacity-achieving linear SC-PIR scheme is lower bounded by
\begin{eqnarray}\label{Eqn_bound}
F\geq\eta^*\cdot(M-1),
\end{eqnarray}
where $\eta^*$ is the optimal value to Problem 1.
\end{Theorem}
\begin{IEEEproof}
Given any capacity-achieving linear SC-PIR scheme with  $\{\alpha_{\mathcal{S}}: \alpha_{\mathcal{S}}\in [0,1], \mathcal{S}\subseteq[1:N]\}$, the sub-packetization
\begin{IEEEeqnarray*}{rCl}
F&\overset{(a)}{=}&\sum_{\mathcal{S}\subseteq[1:N]}F_{\mathcal{S}}\\
&\overset{(b)}{\geq}&\sum_{\substack{\mathcal{S}\subseteq[1:N]\\|\mathcal{S}|=M}}\mathbf{1}(\alpha_{\mathcal{S}}>0)\cdot(M-1) \notag\\
&\overset{(c)}{\geq} &\eta^*\cdot(M-1) , \notag
\end{IEEEeqnarray*}
where $(a)$ follows by \eqref{def:sub-packetization};
$(b)$ holds by Lemma \ref{bound:each subfile};
$(c)$ is due to Lemma \ref{necessary:P1-2} and Remark \ref{remark:equivalent} that $\{\alpha_{\mathcal{S}}: \alpha_{\mathcal{S}}\in [0,1], \mathcal{S}\subseteq[1:N]\}$ of any capacity-achieving SC-PIR scheme must satisfy \eqref{optimize:condition1} and \eqref{optimize:condition2}.
\end{IEEEproof}

\begin{Definition}
The sub-packetization $F$ of a capacity-achieving linear SC-PIR scheme is said to be  optimal if it achieves the equality in \eqref{Eqn_bound}.
\end{Definition}

\section{A Generic Capacity-Achieving Linear SC-PIR Scheme with Optimal Sub-packetization}\label{packets:charact}
In this section, we present a generic construction of capacity-achieving linear SC-PIR schemes with optimal sub-packetization.

\subsection{SF-SC-PIR Schemes Based on Transformation From SF-PIR Schemes to SC-PIR Schemes}
We first introduce a class of SC-PIR schemes in Algorithm \ref{Connection}, which are constructed by a transformation from SF-PIR schemes to SC-PIR schemes, where we term the resultant schemes as \emph{SF-SC-PIR schemes} for convenience.
Actually, the transformation was first characterized in \cite{Mingyue Ji}.

\begin{algorithm}[t]
\caption{Capacity-Achieving SF-SC-PIR Schemes}
\label{Connection}
\begin{algorithmic}[1] 
\REQUIRE A feasible solution $\{\alpha_{\mathcal{S}}\}_{\mathcal{S}\subseteq[1:N],|\mathcal{S}|=M}$  to Problem 1 and a capacity-achieving $(M,K)$ SF-PIR scheme with sub-packetization $F_{\mathrm{SF}}$
\ENSURE Capacity-Achieving SF-SC-PIR Scheme with sub-packetization $F=\eta\cdot F_{\mathrm{SF}}$, where $\eta=\sum\limits_{{\mathcal{S}\subseteq[1:N],|\mathcal{S}|=M}}\textbf{1}(\alpha_{\mathcal{S}}>0)$.
\STATE \textbf{procedure} Storage \label{storage:1}
\FOR {$k\in[1:K]$}\label{for:split}
\STATE \label{partition} Divide $W_k$ into $\{W_{k,\mathcal{S}}:{\mathcal{S}\subseteq[1:N],|\mathcal{S}|=M},\alpha_{\mathcal{S}}>0\}$ such that $H(W_{k,\mathcal{S}})=\alpha_{\mathcal{S}}L$ 
\STATE \label{split:packet} Further divide each $W_{k,\mathcal{S}}$ into $F_{\mathrm{SF}}$ disjointed packets for $\mathcal{S}\subseteq[1:N],|\mathcal{S}|=M,\alpha_{\mathcal{S}}>0$
\ENDFOR \label{end:split}
\FOR {$n\in[1:N]$}
\STATE $Z_n\leftarrow\{W_{k,\mathcal{S}}:k\in[1:K],\mathcal{S}\subseteq[1:N],|\mathcal{S}|=M,\alpha_{\mathcal{S}}>0,n\in\mathcal{S}\}$ \label{storage:SF-SC PIR}
\ENDFOR
\STATE \textbf{end procedure} \label{storage:2}
\STATE \textbf{procedure} Retrieval
\FOR{\textbf{each} $\mathcal{S}\subseteq[1:N],|\mathcal{S}|=M,\alpha_{\mathcal{S}}>0$} \label{retrieval:SC-SF PIR1}
\STATE Employ the $(M,K)$ capacity-achieving SF-PIR scheme\footnotemark[3] independently to retrieve $W_{\theta,\mathcal{S}}$ privately from the $M$ servers in $\mathcal{S}$ and the $K$ packet sets in $\{W_{k,\mathcal{S}}:k\in[1:K]\}$ \label{retrieval:SC-SF PIR}
\ENDFOR \label{retrieval:SC-SF PIR2}
\STATE The user combines $W_{\theta,\mathcal{S}}$ ($\mathcal{S}\subseteq[1:N],|\mathcal{S}|=M,\alpha_{\mathcal{S}}>0$) to recover $W_{\theta}$
\STATE \textbf{end procedure}
\end{algorithmic}
\end{algorithm}

\begin{Theorem}\label{Thm_Transformation}
For any positive integers $N,K,M$  with $M\in[2:N]$, given any feasible solution $\{\alpha_{\mathcal{S}}\}_{\mathcal{S}\subseteq[1:N],|\mathcal{S}|=M}$  to Problem 1 
and any capacity-achieving $(M,K)$ linear SF-PIR scheme with sub-packetization $F_{\mathrm{SF}}$,
the  $(\mu=M/N,N,K)$ linear SF-SC-PIR scheme obtained in Algorithm \ref{Connection} is capacity-achievable  with sub-packetization $F=\eta \cdot F_{\mathrm{SF}}$, where   $\eta=\sum\limits_{{\mathcal{S}\subseteq[1:N],|\mathcal{S}|=M}}\emph{\textbf{1}}(\alpha_{\mathcal{S}}>0)$.
\end{Theorem}
\begin{IEEEproof}
Obviously, the output SF-SC-PIR scheme of Algorithm \ref{Connection} is linear if the input SF-PIR scheme is linear.
Furthermore, by Lines \ref{for:split}-\ref{end:split}, the sub-packetization of the output scheme is $F=\eta\cdot F_{\mathrm{SF}}$.
Consequently, we prove the theorem by showing that the storage design in Algorithm \ref{Connection} is achievable and the SF-SC-PIR  scheme is capacity-achievable while satisfying the constraints of correctness and privacy.

In Line \ref{partition}, each file can be partitioned into $W_{k}=\{W_{k,\mathcal{S}}:{\mathcal{S}\subseteq[1:N],|\mathcal{S}|=M},\alpha_{\mathcal{S}}>0\}$ since
the feasible solution $\{\alpha_{\mathcal{S}}\}_{\mathcal{S}\subseteq[1:N],\mathcal{S}=M}$ that satisfies \eqref{optimize:condition1}  and \eqref{optimize:condition2} has 
\footnotetext[3]{It is not necessary to use different SF-PIR schemes as building blocks since the sub-packetization of resultant SC-PIR scheme can be further reduced by adopting the identical SF-PIR scheme with minimum sub-packetization.}
\begin{IEEEeqnarray}{rCl}
\sum\limits_{\substack{\mathcal{S}\subseteq[1:N]\\|\mathcal{S}|=M,\alpha_{\mathcal{S}}>0}}\alpha_{\mathcal{S}}
&=&\sum\limits_{\substack{\mathcal{S}\subseteq[1:N]\\|\mathcal{S}|=M}}\alpha_{\mathcal{S}} \label{sum:alpha:T}\\
&=&\frac{1}{M}\sum\limits_{\substack{\mathcal{S}\subseteq[1:N]\\|\mathcal{S}|=M}}\alpha_{\mathcal{S}}\sum_{n\in[1:N]}\mathbf{1}(n\in\mathcal{S})\notag\\
&=&\frac{1}{M}\sum_{n\in[1:N]}\sum\limits_{\substack{\mathcal{S}\subseteq[1:N]\\|\mathcal{S}|=M}}\alpha_{\mathcal{S}}\cdot\mathbf{1}(n\in\mathcal{S})\notag\\
&=&\frac{1}{M}\sum_{n\in[1:N]}\sum\limits_{\substack{\mathcal{S}\subseteq[1:N]\\|\mathcal{S}|=M, n\in\mathcal{S}}}\alpha_{\mathcal{S}}\notag\\
&\overset{(a)}{=}&\frac{\mu N}{M}\notag\\
&=&1 \notag,
\end{IEEEeqnarray}
where $(a)$ is due to \eqref{optimize:condition1}.
In Line \ref{storage:SF-SC PIR}, the storage content $Z_n$ at each server is
\begin{IEEEeqnarray}{rCl}
Z_n&=&\mathop\cup\limits_{k\in[1:K]}\mathop\cup\limits_{\substack{\mathcal{S}\subseteq[1:N],|\mathcal{S}|=M\\\alpha_{\mathcal{S}}>0,n\in\mathcal{S}}}W_{k,\mathcal{S}},\quad\forall\, n\in[1:N]. \notag
\end{IEEEeqnarray}
Since all the random variables $W_{k,\mathcal{S}}$ are independent of each other,
by applying $H(W_{k,\mathcal{S}})=\alpha_{\mathcal{S}}L$  and \eqref{optimize:condition1}, we get
\begin{IEEEeqnarray}{rCl}
H(Z_n)&=&KL\sum\limits_{\substack{\mathcal{S}\subseteq[1:N],|\mathcal{S}|=M\\\alpha_{\mathcal{S}}>0,n\in\mathcal{S}}}\alpha_{\mathcal{S}}=\mu KL, \notag
\end{IEEEeqnarray}
which satisfies the storage constraint \eqref{storage constraint}. Thus, the storage design in Algorithm \ref{Connection} is achievable.

Then, we prove the scheme in Algorithm \ref{Connection} to be capacity-achievable.
By Lines \ref{split:packet} and \ref{storage:SF-SC PIR}, for any $\mathcal{S}\subseteq[1:N],|\mathcal{S}|=M, \alpha_{\mathcal{S}}>0$, each $W_{k,\mathcal{S}}$ ($k\in[1:K]$) is partitioned into $F_{\mathrm{SF}}$ packets and is stored at $M$ servers in $\mathcal{S}$.
Thus, in Lines \ref{retrieval:SC-SF PIR1}-\ref{retrieval:SC-SF PIR2}, the non-zero $W_{\theta,\mathcal{S}}$ can be retrieved from servers in $\mathcal{S}$ by employing the capacity-achieving $(M,K)$ SC-PIR schemes independently.
Then,  according to \eqref{def:rate},
the download cost for retrieving $W_{\theta,\mathcal{S}}$ is
\begin{IEEEeqnarray}{c}\notag
D_{\mathcal{S}}=\frac{H(W_{\theta,\mathcal{S}})}{C_{\mathrm{SF}}}=\left( 1+\frac{1}{M}+\ldots+\frac{1}{M^{K-1}} \right)\alpha_{\mathcal{S}}L,
\end{IEEEeqnarray}
where $C_{\mathrm{SF}}={(1+1/M+\ldots+1/M^{K-1})}^{-1}$ is the capacity of $(M,K)$ SF-PIR scheme.
Therefore,  the rate for retrieving $W_{\theta}$ is
\begin{IEEEeqnarray}{c}\notag
R=\frac{L}{D}=\frac{L}{\sum\limits_{\substack{\mathcal{S}\subseteq[1:N]\\|\mathcal{S}|=M,\alpha_{\mathcal{S}}>0}} D_{\mathcal{S}}}=\left( 1+\frac{1}{M}+\ldots+\frac{1}{M^{K-1}} \right)^{-1},
\end{IEEEeqnarray}
which achieves the capacity of SC-PIR in \eqref{optimal download}.

The user can recover $W_{\theta}$ by combining all non-zero $W_{\theta,\mathcal{S}}$, where privacy is guaranteed because the SF-PIR scheme satisfying the constraint of privacy is independently employed to download the desired packet sets.
\end{IEEEproof}

We know from Theorem \ref{Thm_Transformation} that any parameters $\{\alpha_{\mathcal{S}}: \alpha_{\mathcal{S}}\in [0,1], \mathcal{S}\subseteq[1:N]\}$ satisfying \eqref{optimize:condition1} and \eqref{optimize:condition2} result in a storage design of a capacity-achieving SC-PIR scheme.
Thus,  we have the following corollary according to Remark \ref{remark:equivalent}.
%
\begin{Corollary}\label{remark:storage:conditions}
Given any parameters $\{\alpha_{\mathcal{S}}: \alpha_{\mathcal{S}}\in [0,1], \mathcal{S}\subseteq[1:N]\}$, P1 and P2 are necessary and sufficient conditions for the storage design of a capacity-achieving SC-PIR scheme.
\end{Corollary}


\subsection{Capacity-Achieving Linear SC-PIR Schemes with Optimal Sub-packetization}
From Algorithm \ref{Connection}, we can construct a capacity-achieving SC-PIR scheme by using any feasible solution to Problem 1 and any capacity-achieving SF-PIR scheme as a building block.
Such capacity-achieving $(M,K)$ SF-PIR schemes have been found in \cite{Sun replicated,Sun optimal,Tian and Sun}. If the SF-PIR scheme with sub-packetization $M^{K-1}$ \cite{Sun optimal} is employed, then we can obtain a capacity-achieving SC-PIR scheme with sub-packetization $\eta\cdot M^{K-1}$, which has identical download cost across all random realizations of queries. Whereas, if the scheme with sub-packetization $M-1$ in \cite{Tian and Sun} is adopted, the asymmetry of download cost over all realizations of queries will be inherited by the resultant SC-PIR scheme so that the sub-packetization is reduced to $\eta\cdot (M-1)$.
In particular, when any optimal solution to Problem 1 is further employed, a capacity-achieving SF-SC-PIR scheme with sub-packetization $\eta^{*}\cdot (M-1)$ can be obtained.

For the sake of completeness, we summarize the scheme of \cite{Tian and Sun} in Algorithm \ref{SF-PIR Scheme}, where $\mathcal{Q}$ is defined as
\begin{IEEEeqnarray}{c}\notag
\mathcal{Q}\triangleq\left\{(q_1,\ldots,q_{K})\in[0:M-1]^{K}\right\}.
\end{IEEEeqnarray}

\begin{algorithm}[http]
\caption{Capacity-Achieving $(M,K)$ Linear SF-PIR Scheme with Sub-packetization $M-1$}
\label{SF-PIR Scheme}
\begin{algorithmic}[1] 
\STATE Relabel the indices of the $M$ servers as $0,1,\ldots,M-1$.
\STATE \label{dummy packets} For each $k\in[1:K]$, file $W_k$ is uniformly partitioned into $M-1$ disjoint packets $W_{k,0},W_{k,1},\ldots,W_{k,M-2}$. For easy of exploration, each file $W_{k}$ is appended a dummy packet $W_{k,M-1}\triangleq\textbf{0}$, i.e.,
\begin{IEEEeqnarray}{c}\notag
W_{k}=\left( W_{k,0},\ldots,W_{k,M-2},\textbf{0} \right),\quad\forall\, k\in[1:K].
\end{IEEEeqnarray}
\STATE Select a vector from the set $\mathcal{Q}$ \emph{independently and uniformly}:
\begin{IEEEeqnarray}{c}\notag
\mathbf{q}=(q_1,\ldots,q_{\theta-1},q_{\theta},q_{\theta+1},\ldots,q_{K}).
\end{IEEEeqnarray}
\STATE \emph{Query Phase: }Based on the vector $\mathbf{q}$, the user constructs a query sent to server $m$ as
\begin{IEEEeqnarray}{c}\notag
Q_{m}^{[\theta]}=(q_1,\ldots,q_{\theta-1},(q_{\theta}+m)_{M},q_{\theta+1},\ldots,q_{K}), \quad\forall\, m\in[0:M-1].
\end{IEEEeqnarray}
\STATE \label{answer:SF-PIR}\emph{Answer Phase: }After receiving the query $Q_{m}^{[\theta]}$, the answer at server $m$ is
\begin{IEEEeqnarray}{c}\notag
A_{m}^{[\theta]}=\left\{\begin{array}{@{}ll}
                                \mathrm{NULL},   &\mathrm{if}~ Q_{m}^{[\theta]}=(M-1,\ldots,M-1)  \\
                                 \sum\limits_{i\in[1:K]\backslash\{\theta\}}W_{i,q_{i}}+W_{\theta,(q_{\theta}+m)_{M}},&\mathrm{else}
                                \end{array}
      \right.,\quad\forall\, m\in[0:M-1],\label{server answer}
\end{IEEEeqnarray}
where the value $\mathrm{NULL}$ indicates that the server keeps silence.
\STATE \emph{Decoding Phase: }Decode file $W_{\theta}=\left( W_{\theta,0},\ldots,W_{\theta,M-2} \right)$ from the answers $A_{0}^{[\theta]},\ldots,A_{M-1}^{[\theta]}$
\end{algorithmic}
\end{algorithm}

Note that the dummy packets in Line \ref{dummy packets} are not stored by the servers at all. Let $\Delta=\sum_{i\in[1:K]\backslash\{\theta\}}W_{i,q_{i}}$,
then the user can decode file $W_{k}=\left( W_{k,0},\ldots,W_{k,M-2} \right)$ from the answers $(A_{0}^{[\theta]},\ldots,A_{M-1}^{[\theta]})$ in Line \ref{answer:SF-PIR} because of $(A_{0}^{[\theta]},\ldots,A_{M-1}^{[\theta]})=\left(\Delta+W_{\theta,q_{\theta}},\ldots,\Delta+W_{\theta,M-2},\Delta+\textbf{0},\Delta+W_{\theta,0},\ldots,\Delta+W_{\theta,q_{\theta}-1}\right)$.

The following result is immediate by Theorems \ref{arbitra:part} and \ref{Thm_Transformation}.

\begin{Theorem}\label{arbitra:part2}
For any positive integers $N,K,M$  with $M\in[2:N]$, given any optimal solution $\{\alpha^{*}_{\mathcal{S}}\}_{\mathcal{S}\subseteq[1:N],|\mathcal{S}|=M}$  to Problem 1 and the capacity-achieving $(M,K)$ linear SF-PIR scheme in Algorithm \ref{SF-PIR Scheme}, Algorithm \ref{Connection} outputs  a capacity-achieving $(\mu=M/N,N,K)$ linear SC-PIR scheme with sub-packetization $F^*=\eta^*\cdot(M-1)$, where $\eta^*=\sum\limits_{{\mathcal{S}\subseteq[1:N],|\mathcal{S}|=M}}\emph{\textbf{1}}(\alpha^*_{\mathcal{S}}>0)$.
Particularly, the sub-packetization $F^*$ is optimal among all capacity-achieving linear SC-PIR schemes.
\end{Theorem}

\section{Storage Design Array}\label{SDA:def}
According to Theorem \ref{arbitra:part2}, in terms of designing capacity-achieving linear SC-PIR schemes with optimal sub-packetization, it is crucial  to solve the optimization problem in Problem 1.
However, it is not easy because of the involved indicator functions. Thus, in this section, we dedicate to construct concrete  capacity-achieving linear schemes with low sub-packetization by finding  sub-optimal solutions to Problem 1.

For clarity, we first introduce Storage Design Array (SDA) to construct feasible solutions of Problem 1.

\begin{Definition}[Storage Design Array (SDA)] For any positive integers $N,M$  with $M\in[1:N]$,
an $(N,M)$ storage design array is an array of size $N\times \frac{N}{\gcd(N,M)}$ with  each entry being either ``$\ast$" or ``$\mathrm{NULL}$" that satisfies
\begin{itemize}
  \item [S1.] Each column has  $M$ ``$*$"s;
  \item [S2.] Each row has $\frac{M}{\gcd(N,M)}$ ``$*$"s.
\end{itemize}
\end{Definition}

\begin{Definition}[Number of Distinct Columns of SDA]\label{def:eta:P}
Let $\mathbf{P}=[p_{i,j}]_{N\times \frac{N}{\gcd(N,M)}}$ be an $(N,M)$ SDA. For each $j\in[1:\frac{N}{\gcd(N,M)}]$, let $\mathcal{S}_j$ be the set of row indices  corresponding to ``$*$"s in column $j$, i.e.,
\begin{IEEEeqnarray}{c}
\mathcal{S}_{j}\triangleq\{i\in[1:N]:p_{i,j}=*\},\quad\forall\, j\in\left[1:\frac{N}{\gcd(N,M)}\right].\label{def:Sj}
\end{IEEEeqnarray}
We denote $\eta_{\mathbf{P}}$ as the number of distinct columns in $\mathbf{P}$, i.e.,
\begin{IEEEeqnarray}{c}
\eta_\mathbf{P}\triangleq\left|\left\{\mathcal{S}_j:j\in\left[1:\frac{N}{\gcd(N,M)}\right]\right\}\right|.\label{def:etaP}
\end{IEEEeqnarray}
Let $\{\mathcal{S}_{i_l}\}_{l=1}^{\eta_\mathbf{P}}$ be the $\eta_{\mathbf{P}}$ distinct ones in $\{\mathcal{S}_j\}_{j=1}^{\frac{N}{\gcd(N,M)}}$ and $s_{l}$  be the occurrence that $\mathcal{S}_{i_l}$ appears in $\{\mathcal{S}_j\}_{j=1}^{\frac{N}{\gcd(N,M)}}$ for $l\in[1:\eta_{\mathbf{P}}]$, i.e.,
\begin{IEEEeqnarray}{c}\label{S:l}
s_l\triangleq\left|\left\{j\in\left[1:\frac{N}{\gcd(N,M)}\right]:\mathcal{S}_{j}=\mathcal{S}_{i_l}\right\}\right|, \quad\forall\, l\in[1:\eta_{\mathbf{P}}].
\end{IEEEeqnarray}
\end{Definition}

\begin{Example}\label{SDA_def}
An $(N=9,M=4)$ SDA $\mathbf{P}$ and another $(N=11,M=5)$ SDA $\mathbf{P}'$ can be written as follows, respectively.
\begin{IEEEeqnarray}{c}\notag
\mathbf{P}=
    \left[
      \begin{array}{ccccccccc}
      * & * & * & *  &   &   &  &  & \\
      * & * & * & *  &   &   &  &  & \\
      * & * & * & *  &   &   &  &  & \\
      * & * & *  & *  &   &   &  &  & \\
       &  &   &   &  * &  * & * & * & \\
       &  &   &   &  * &  * & * &  & *\\
       &  &   &   &  * &  * &  & * & *\\
       &  &   &   &  * &   & * & * &* \\
       &  &   &   &   &  * & * & * & *\\
      \end{array}
    \right]_{9\times 9},\quad
\mathbf{P}'=
\left[
\begin{array}{ccccccccccc}
  * & * & * & * & * &   &   &   &   &   &   \\
  * & * & * & * & * &   &   &   &   &   &   \\
  * & * & * & * & * &   &   &   &   &   &   \\
  * & * & * & * & * &   &   &   &   &   &   \\
  * & * & * & * & * &   &   &   &   &   &   \\
    &   &   &   &   & * & * & * & * & * &   \\
    &   &   &   &   & * & * & * & * &   & * \\
    &   &   &   &   & * & * & * &   & * & * \\
    &   &   &   &   & * & * &   & * & * & * \\
    &   &   &   &   & * &   & * & * & * & * \\
    &   &   &   &   &   & * & * & * & * & *
\end{array}
\right]_{11\times 11}.
\end{IEEEeqnarray}
In fact, the SDAs $\mathbf{P}$ and $\mathbf{P}'$ are constructed by greedy Algorithm \ref{G-SDA alg}, which will be illustrated in Section \ref{G-SDA} in detail.
Apparently, there are $\eta_{\mathbf{P}}=6$ distinct columns in $\mathbf{P}$ (i.e., columns $1,5,6,7,8$ and $9$) with $\{s_1=4,s_2=1,s_3=1,s_4=1,s_5=1,s_6=1\}$ and $\eta_{\mathbf{P}'}=7$ distinct columns in $\mathbf{P}'$ (i.e., columns $1,6,7,8,9,10$ and $11$) with $\{s_1=5,s_2=1,s_3=1,s_4=1,s_5=1,s_6=1,s_7=1\}$.
\end{Example} 



\begin{Lemma}\label{lemma:solution}
Given $N$ and $M\in[1:N]$, any $(N,M)$ SDA $\mathbf{P}$ is associated to a set of parameters $\{\alpha_{\mathcal{S}}\}_{\mathcal{S}\subseteq[1:N],|\mathcal{S}|=M}$ that is a feasible solution to Problem 1. 
\end{Lemma}
\begin{IEEEproof}
Notice from S1 and \eqref{def:Sj} that $|\mathcal{S}_{i_l}|=M$ for all $l\in[1:\eta_{\mathbf{P}}]$.
Thus, we can obtain a set of parameters $\{\alpha_{\mathcal{S}}\}_{\mathcal{S}\subseteq[1:N],|\mathcal{S}|=M}$,
\begin{IEEEeqnarray}{c}\label{alpha:S:P}
\alpha_{\mathcal{S}}=\left\{\begin{array}{@{}ll}
s_l \cdot\frac{\gcd(N,M)}{N},&\mathrm{if}~\mathcal{S}=\mathcal{S}_{i_l}~\textnormal{for some}~l\in[1:\eta_{\mathbf{P}}]  \\
                              0, &\mathrm{otherwise}
                            \end{array}
\right.,
\end{IEEEeqnarray}
where $s_l$ is defined in \eqref{S:l}. 
Then, for any $n\in[1:N]$,
\begin{IEEEeqnarray}{rCl}
\sum\limits_{\substack{\mathcal{S}\subseteq[1:N]\\|\mathcal{S}|=M,n\in\mathcal{S}}}\alpha_{\mathcal{S}}&=&\sum\limits_{\substack{l\in[1:\eta_{\mathbf{P}}]\\n\in\mathcal{S}_{i_l}}}\alpha_{\mathcal{S}_{i_l}} \notag\\
&=&\frac{\gcd(N,M)}{N}\cdot\sum\limits_{\substack{l\in[1:\eta_{\mathbf{P}}]\\n\in\mathcal{S}_{i_l}}}s_{l} \notag\\
&\overset{(a)}{=}&\frac{\gcd(N,M)}{N}\cdot\sum\limits_{\substack{j\in[1:\frac{N}{\gcd(N,M)}]}}\mathbf{1}(p_{n,j}=*)\notag\\
&\overset{(b)}{=}&\frac{\gcd(N,M)}{N}\cdot \frac{M}{\gcd(N,M)}\notag\\
&=&\mu, \notag
\end{IEEEeqnarray}
where $(a)$ follows from \eqref{def:Sj} and  \eqref{S:l}, and $(b)$ is due to S2. That is,  the parameters $\{\alpha_{\mathcal{S}}\}_{\mathcal{S}\subseteq[1:N],|\mathcal{S}|=M}$ satisfy \eqref{optimize:condition1} and \eqref{optimize:condition2}, and thus are  feasible for Problem 1. 
\end{IEEEproof}
\vspace{0.8mm}
Obviously, taking the feasible solution $\{\alpha_{\mathcal{S}}\}_{\mathcal{S}\subseteq[1:N],|\mathcal{S}|=M}$ and the $(M,K)$ SF-PIR scheme in Algorithm \ref{SF-PIR Scheme} as inputs of Algorithm \ref{Connection}, one can obtain a storage design scheme by Lines \ref{storage:1}-\ref{storage:2} in Algorithm \ref{Connection} and a capacity-achieving SF-SC-PIR scheme with sub-packetization $\eta_{\mathbf{P}}\cdot(M-1)$ by Theorem \ref{Thm_Transformation}, where $\eta_{\mathbf{P}}=\sum\limits_{{\mathcal{S}\subseteq[1:N],|\mathcal{S}|=M}}{\textbf{1}}(\alpha_{\mathcal{S}}>0)$ by \eqref{alpha:S:P}.

\begin{Theorem}\label{relation: SDA and Problem2}
Given any positive integers $N,K,M$  with $M\in[2:N]$  and any $(N,M)$ SDA $\mathbf{P}$, there is a  capacity-achieving $(\mu=M/N,N,K)$ linear SC-PIR scheme  with sub-packetization $\eta_\mathbf{P}\cdot(M-1)$.
\end{Theorem}

\begin{Example}\label{Example11}
For the $(N=9,M=4)$ SDA $\mathbf{P}$ in Example \ref{SDA_def}, set
\begin{IEEEeqnarray}{c}\notag
\alpha_{\{1,2,3,4\}}=\frac{4}{9},~\alpha_{\{5,6,7,8\}}=\alpha_{\{5,6,7,9\}}=\alpha_{\{5,6,8,9\}}=\alpha_{\{5,7,8,9\}}=\alpha_{\{6,7,8,9\}}=\frac{1}{9},
\end{IEEEeqnarray}
and all the other $\alpha_{\mathcal{S}}$ to be zeros. It is easy to see that $\{\alpha_{\mathcal{S}}\}_{\mathcal{S}\subseteq[1:9],|\mathcal{S}|=4}$ is a feasible solution of Problem 1 with $\sum\limits_{{\mathcal{S}\subseteq[1:9],|\mathcal{S}|=4}}\textbf{1}(\alpha_{\mathcal{S}}>0)=\eta_{\mathbf{P}}=6$.
Then, we can generate a capacity-achieving $(\mu={4}/{9},N=9,K)$ linear SC-PIR scheme with sub-packetization $18$. Similarly, the $(N=11,M=5)$ SDA $\mathbf{P}'$ in Example \ref{SDA_def} is associated to a capacity-achieving $(\mu={5}/{11},N=11,K)$ linear SC-PIR scheme with sub-packetization $28$.
\end{Example}

\section{Equal-Size Capacity-Achieving Linear SC-PIR Schemes}\label{equal-size packets}
Recall that the setup in Section \ref{system model} allows us to partition each file into unequal-size packets.
Actually, the equal-size partition of the files is one of the most important cases in practice, which is also considered by the previous capacity-achieving SC-PIR schemes \cite{Tandon Coded caching,Attia SC-PIR,Mingyue Ji} and SF-PIR schemes \cite{Sun replicated,Sun optimal,Tian and Sun} in the storage phase.
Thus, we first focus on capacity-achieving SC-PIR/SF-PIR schemes with small sub-packetization by imposing the assumption of equal-size packets, i.e.,
\begin{IEEEeqnarray}{c}
H(W_{k,1})=H(W_{k,2})=\ldots=H(W_{k,F})=\frac{L}{F},\quad \forall\, k\in[1:K].\label{equal:size}
\end{IEEEeqnarray}
For simplicity, we will refer to the sub-packetization of a scheme satisfying  \eqref{equal:size} as \emph{equal-size sub-packetization}.  In particular, we characterize the optimal equal-size sub-packetization of all capacity-achieving linear SF-SC-PIR schemes in the following theorem.

\begin{Theorem}\label{thm:equalsize}
Given any $(\mu,N,K)$ SC-PIR system with $M=\mu N\in[2:N]$, the optimal equal-size sub-packetization of all  capacity-achieving linear SF-SC-PIR schemes is given by $\frac{N(M-1)}{\gcd(N,M)}$.
\end{Theorem}

The theorem will be proved by constructing an SDA-based SF-SC-PIR scheme with equal-size sub-packetization and  showing the optimality of its sub-packetization separately.

\subsection{SDA-Based SF-SC-PIR Schemes with Equal-size Sub-packetization}\label{Equal-SDA}
In this subsection, we construct an $(N,M)$ SDA $\mathbf{P}$ with all columns being distinct, i.e., $\eta_{\mathbf{P}}=\frac{N}{\gcd(N,M)}$. Later, it will be shown that such SDA $\mathbf{P}$ is associated to a capacity-achieving SF-SC-PIR scheme with equal-size sub-packetization $\frac{N(M-1)}{\gcd(N,M)}$.


%
%

Before that, a simple example is presented.
\begin{Example}\label{comparision}
For $N=12$ and $M=5$, an SDA with all columns being distinct is given by
\begin{IEEEeqnarray}{c}\notag
\mathbf{P}=
    \left[
      \begin{array}{cccccccccccc}
      * &   & * &   & * &   &   & * &   & * &   &   \\
      * &   & * &   &   & * &   & * &   & * &   &   \\
      * &   & * &   &   & * &   & * &   &   & * &   \\
      * &   &   & * &   & * &   & * &   &   & * &   \\
      * &   &   & * &   & * &   &   & * &   & * &   \\
        & * &   & * &   & * &   &   & * &   & * &   \\
        & * &   & * &   &   & * &   & * &   & * &   \\
        & * &   & * &   &   & * &   & * &   &   & * \\
        & * &   &   & * &   & * &   & * &   &   & * \\
        & * &   &   & * &   & * &   &   & * &   & * \\
        &   & * &   & * &   & * &   &   & * &   & * \\
        &   & * &   & * &   &   & * &   & * &   & *
      \end{array}
    \right]_{12\times 12}.
\end{IEEEeqnarray}
Then it corresponds to a set of non-zero and equal-size parameters
\begin{IEEEeqnarray}{rCl}
&&\alpha_{\{1,2,3,4,5\}}=\alpha_{\{6,7,8,9,10\}}=\alpha_{\{11,12,1,2,3\}}=\alpha_{\{4,5,6,7,8\}}=\alpha_{\{9,10,11,12,1\}}=\alpha_{\{2,3,4,5,6\}}\\
&=&\alpha_{\{7,8,9,10,11\}}=\alpha_{\{12,1,2,3,4\}}=\alpha_{\{5,6,7,8,9\}}=\alpha_{\{10,11,12,1,2\}}=\alpha_{\{3,4,5,6,7\}}=\alpha_{\{8,9,10,11,12\}}=\frac{1}{12}. \notag
\end{IEEEeqnarray}
By employing these parameters and the $(M=5,K)$ SF-PIR scheme in  Algorithm \ref{SF-PIR Scheme} as inputs of Algorithm \ref{Connection},   a capacity-achieving SF-SC-PIR scheme is obtained, where each packet has equal size $\frac{1}{48}L$ and  sub-packetization $48$.
\end{Example}

Formally, an $(N,M)$ SDA $\mathbf{P}=[p_{i,j}]_{N\times\frac{N}{\gcd(N,M)}}$ satisfying $\eta_{\mathbf{P}}=\frac{N}{\gcd(N,M)}$ is constructed as
\begin{IEEEeqnarray}{c}\label{equal_size:SDA}
p_{i,j}=\left\{\begin{array}{@{}ll}
                 *, &\mathrm{if}~i\in\mathcal{S}_j \\
                \mathrm{NULL}, &\mathrm{if}~i\notin\mathcal{S}_j
               \end{array}
\right.,
\end{IEEEeqnarray}
where
\begin{IEEEeqnarray}{c}\label{Def:storage set}
\mathcal{S}_{j}\triangleq\big([0:M-1]+(j-1)\cdot M\big)_N+1,\quad\forall\, j\in\left[1:\frac{N}{\gcd(N,M)}\right].
\end{IEEEeqnarray}

It is easy to check that the array $\mathbf{P}$ is an $(N,M)$ SDA satisfying $\eta_{\mathbf{P}}=\frac{N}{\gcd(N,M)}$, by the following three facts from $\mathcal{S}_{j}$ $(j\in\big[1:\frac{N}{\gcd(N,M)}\big])$:
\begin{itemize}
  \item[F1.] For each $j\in\big[1:\frac{N}{\gcd(N,M)}\big]$, set $\mathcal{S}_j$ is of size $M$, i.e., $|\mathcal{S}_j|=M$ for all $j\in [1:\frac{N}{\gcd(N,M)}]$;
  \item[F2.] All the sets in \eqref{Def:storage set} are distinct, i.e., $\mathcal{S}_i\neq\mathcal{S}_j$ for any $i\ne j\in [1:\frac{N}{\gcd(N,M)}]$;
  \item[F3.] For any given $n\in[1:N]$, $n$ exactly occurs in $\frac{M}{\gcd(N,M)}$ different sets in \eqref{Def:storage set}, i.e., $|\{j\in[1:\frac{N}{\gcd(N,M)}]:n\in\mathcal{S}_j\}|=\frac{M}{\gcd(N,M)}$ for all $n\in[1:N]$.
\end{itemize}

By \eqref{alpha:S:P}, $\alpha_{\mathcal{S}}=\frac{\gcd(N,M)}{N}$ if $\mathcal{S}=\mathcal{S}_j$ for some $j\in[1:\frac{N}{\gcd(N,M)}]$, and $\alpha_{\mathcal{S}}=0$ otherwise.
Accordingly,  all the non-zero $\{W_{k,\mathcal{S}}:k\in[1:K],\mathcal{S}\subseteq[1:N],|\mathcal{S}|=M,\alpha_{\mathcal{S}}>0\}$ are of equal size and each is  partitioned into $M-1$ equal-size packets by Algorithms \ref{Connection}-\ref{SF-PIR Scheme}. Therefore, its capacity-achieving SF-SC-PIR  scheme has equal-size sub-packetization $\frac{N(M-1)}{\gcd(N,M)}$.


\subsection{Optimality of Equal-Size Sub-packetization}\label{optimality:equal packets}
Recall from Algorithm \ref{Connection} that any feasible solution $\{\alpha_{\mathcal{S}}\}_{\mathcal{S}\subseteq[1:N],|\mathcal{S}|=M}$ to Problem 1 can support a capacity-achieving $(\mu=M/N,N,K)$ linear SF-SC-PIR scheme by employing any specific capacity-achieving $(M,K)$ linear SF-PIR scheme as a building block.
According to Line 4 in Algorithm \ref{Connection} and \eqref{equal:size}, each $W_{k,\mathcal{S}}$ of size $\alpha_{\mathcal{S}}>0$ is partitioned into  $F_{\mathrm{SF}}$ equal-sized disjoint packets.
Thus, to design a linear SF-SC-PIR scheme with equal-size sub-packetization, it is necessary that $\alpha_{\mathcal{S}}>0$ is a constant.
\begin{Lemma}\label{lem:equal-szie}
Given any $(\mu,N,K)$ SC-PIR system with $M=\mu N\in[2:N]$ and $\{\alpha_{\mathcal{S}}\}_{\mathcal{S}\subseteq[1:N]}$, the storage design of any capacity-achieving linear SF-SC-PIR scheme with equal-size sub-packetization must satisfy:
\begin{enumerate}
  \item[P6.] The equal-size partition storage is adopted, i.e., all the non-zero $\alpha_{\mathcal{S}}$ has the same value.
\end{enumerate}
\end{Lemma}

From Theorem \ref{arbitra:part}, the equal-size sub-packetization of any SF-SC-PIR scheme is no less than $\eta_e^*\cdot(M-1)$, where $\eta_e^*$ is the optimal value of the following problem by Lemmas \ref{necessary:P1-2} and \ref{lem:equal-szie}.
\begin{Definition}
Given any positive integers $N$ and $M=\mu N\in[2:N]$, \textbf{Problem 2} is defined as
\begin{IEEEeqnarray}{ccLl}
(\{\alpha_{\mathcal{S}}^{*}\}_{\mathcal{S}\subseteq[1:N],|\mathcal{S}|=M},\Delta^*)=\;& \arg\min &\quad \sum\limits_{\substack{\mathcal{S}\subseteq[1:N]\\|\mathcal{S}|=M}}\mathbf{1}(\alpha_{\mathcal{S}}>0)&\notag  \\
&{s.t.} &\quad \sum\limits_{\substack{\mathcal{S}\subseteq[1:N]\\|\mathcal{S}|=M, n\in\mathcal{S}}}\alpha_{\mathcal{S}}=\mu,\quad& \forall\,n\in[1:N]  \label{opti:equal1}\\
&&\quad\alpha_{\mathcal{S}}\in\{\Delta,0\},&\forall\,\mathcal{S}\subseteq[1:N],|\mathcal{S}|=M \label{opti:equal2}\\
&&\quad 0\leq \Delta\leq 1
\label{opti:equal3}
\end{IEEEeqnarray}
where  $(\{\alpha_{\mathcal{S}}^{*}\}_{\mathcal{S}\subseteq[1:N],|\mathcal{S}|=M},\Delta^*)$ is called the optimal solution to Problem 2 and $\eta^*_e=\sum\limits_{\substack{\mathcal{S}\subseteq[1:N],|\mathcal{S}|=M}}\mathbf{1}(\alpha_{\mathcal{S}}^*>0)$ is
called the optimal value of Problem 2.
\end{Definition}

Thus, we just need to prove that the optimal value $\eta_{e}^{*}$ of Problem 2 satisfies $\eta_{e}^{*}\geq\frac{N}{\gcd(N,M)}$.
From Problem 2,
\begin{IEEEeqnarray}{rCl}
1&\overset{(a)}{=}&\sum\limits_{\substack{\mathcal{S}\subseteq[1:N]\\|\mathcal{S}|=M}}\alpha_{\mathcal{S}}^{*} \notag\\
&=&\Delta^*\cdot\sum\limits_{\substack{\mathcal{S}\subseteq[1:N]\\|\mathcal{S}|=M}}\textbf{1}(\alpha_{\mathcal{S}}^{*}>0) \notag\\
&=&\Delta^*\cdot \eta_{e}^{*}, \notag
\end{IEEEeqnarray}
where $(a)$ follows by \eqref{sum:alpha:T} and \eqref{opti:equal1}-\eqref{opti:equal3}. Thus, we have $\Delta^*=\frac{1}{\eta_{e}^{*}}$.

By \eqref{opti:equal1}, the storage constraint of any server $n$ is
\begin{IEEEeqnarray}{rCl}
\mu&=&\sum\limits_{\substack{\mathcal{S}\subseteq[1:N]\\|\mathcal{S}|=M,n\in\mathcal{S}}}\alpha_{\mathcal{S}}^{*} \notag\\
&=&\Delta^*\cdot\sum\limits_{\substack{\mathcal{S}\subseteq[1:N]\\|\mathcal{S}|=M,n\in\mathcal{S}}}\mathbf{1}(\alpha_{\mathcal{S}}^{*}>0) \notag\\
&=&\frac{1}{\eta_{e}^{*}}\nu. \notag
\end{IEEEeqnarray}
Note that $\nu=\sum\limits_{\mathcal{S}\subseteq[1:N],\\|\mathcal{S}|=M,n\in\mathcal{S}}\mathbf{1}(\alpha_{\mathcal{S}}^{*}>0)$ is an integer. Then, the above equation indicates that
\begin{IEEEeqnarray}{c}
 \nu=\mu\cdot\eta_{e}^{*}=\frac{M}{N}\cdot\eta_{e}^{*} \notag
 \end{IEEEeqnarray}
 is an integer.
  Therefore, $\eta_{e}^{*}$ must be greater than or equal to $\frac{N}{\gcd(N,M)}$, which completes the proof of Theorem \ref{thm:equalsize}.

\section{Capacity-Achieving Linear SC-PIR Schemes with Lower Sub-packetization}\label{schemes SDA}

The sub-packetization reflects the implementation complexity of a scheme, specifically in a PIR system, low sub-packetization achieves low complexity \cite{Tian and Sun}. In order to further reduce sub-packetization,
we allow unequal-size packets in this section.
Notably, 
unequal-size packets are usually unavoidable in such SC-PIR \cite{Tandon Coded caching,Attia SC-PIR},  since the memory-sharing technique typically results in schemes with unequal-size packets \cite{Attia SC-PIR,Maddah-Ali}. Particularly, memory-sharing is often  used to achieve the capacity for any storage $M\in[1,N]$ by resorting to the discrete points with $M=1,2,\ldots,N$.



Next, we first present an example to illustrate that allowing unequal-size packets can further decrease sub-packetization of capacity-achieving SC-PIR schemes.
\begin{Example}\label{Example1}
An $(N=12,M=5)$ SDA can be also constructed by the form of
\begin{IEEEeqnarray}{c}\notag
\mathbf{P}=
    \left[
      \begin{array}{cccccccccccc}
        * & * & * & * & * &  &  &  &  &  &  &  \\
        * & * & * & * & * &  &  &  &  &  &  &  \\
        * & * & * & * & * &  &  &  &  &  &  &  \\
        * & * & * & * & * &  &  &  &  &  &  &  \\
        * & * & * & * & * &  &  &  &  &  &  &  \\
        &  &  &  &  & * & * & * & * & * &  &  \\
        &  &  &  &  & * & * & * & * &  & * &  \\
        &  &  &  &  & * & * & * & * &  &  & * \\
        &  &  &  &  & * & * &  &  & * & * & * \\
        &  &  &  &  & * & * &  &  & * & * & * \\
        &  &  &  &  &  &  & * & * & * & * & * \\
        &  &  &  &  &  &  & * & * & * & * & *
      \end{array}
    \right]_{12\times 12}.
\end{IEEEeqnarray}
There are $\eta_{\mathbf{P}}=6$ distinct columns in $\mathbf{P}$, i.e., columns $1,6,8,10,11$ and $12$.
By Theorem \ref{relation: SDA and Problem2}, the SDA can be used for constructing a capacity-achieving $(\mu={5}/{12},N=12,K)$ linear SC-PIR scheme with sub-packetization $24$, which is smaller than 48,  the optimal equal-size sub-packetization as illustrated in Example \ref{comparision}.
\end{Example} 

Based on Theorem \ref{relation: SDA and Problem2}, we wish to construct an SDA $\mathbf{P}$ with $\eta_{\mathbf{P}}$ as low as possible for further reducing sub-packetization.

\subsection{Greedy Construction of Storage Design Arrays}\label{G-SDA}
In this subsection, we propose a greedy construction of $(N,M)$ SDA $\mathbf{P}$ for any $N$ and $M\in[1:N]$.
By convenience, for any positive integers $n,m$, we use $[*]_{n\times m}$ to denote an array of size $n\times m$ with all the entries being ``$*$"s.

Clearly, when $\gcd(N,M)>1$,  an $(N,M)$ SDA $\mathbf{P}$  can be yielded by
\begin{IEEEeqnarray}{c}\label{SDA:gcd}
\mathbf{P}=\left[\begin{array}{c}
                       \mathbf{P}'\\
                       \vdots\\
                       \mathbf{P}'
                   \end{array}\right]_{N\times\frac{N}{\gcd(N,M)}}
\hspace{-2.37cm}
\begin{array}{c}
\multirow{3}{*}{$\left.\begin{array}{c}\\ \\ \\\end{array}\right\}\gcd(N,M)$}\\
\\
\\
\end{array},
\end{IEEEeqnarray}
where $\mathbf{P}'$ is an $(\frac{N}{\gcd(N,M)},\frac{M}{\gcd(N,M)})$ SDA of size $\frac{N}{\gcd(N,M)}\times\frac{N}{\gcd(N,M)}$. That is, $\mathbf{P}$ is generated by repeating $\mathbf{P}'$ $\gcd(N,M)$ times.
Hereafter, we only concentrate on the construction of $(N,M)$ SDA for $\gcd(N,M)=1$. 

Notice that we aim to construct $(N,M)$ SDA $\mathbf{P}$ with $\eta_{\mathbf{P}}$ as small as possible for $\gcd(N,M)=1$. Intuitively, by the definition of $\eta_{\mathbf{P}}$ in \eqref{def:etaP}, the columns of SDA should be repeated as much as possible to reduce $\eta_{\mathbf{P}}$, i.e., $s_l$ should be as big as possible for every $l\in[1:\eta_{\mathbf{P}}]$.
However, $s_l$ is upper bounded by $\min\{M,N-M\}$ since
\begin{itemize}
\item $s_l\leq M$ follows from S2 directly.
\item If $s_l>N-M$, i.e., some column is repeated more than $N-M$ times, without loss of  generality, assume that the first $M$ rows and first $m$ $(m>N-M)$ columns of $\mathbf{P}$
form array $[*]_{M\times m}$. Then by S1, the first $m$ entries of  the $(M+1)$-th row are $\mathrm{NULL}$ and thus  the $(M+1)$-th row has at most $N-m<M$ ``$*$''s, contradicting S2.
\end{itemize}
Based on the above fact that each column in SDA is repeated at most $\min\{M,N-M\}$ times,
 Algorithm \ref{G-SDA alg} is proposed to recursively construct $(N,M)$ SDA for $\gcd(N,M)=1$ as follows:
  \begin{enumerate}
    \item \textbf{Case 1 (Lines \ref{line:if:2}-\ref{P:form}):} If $N-M\geq M$ (i.e., $N\geq 2M$),  $\min\{M,N-M\}=M$. Thus,
greedily generate $[*]_{M\times M}$ and then proceed an $(N-M,M)$ SDA.
     \item \textbf{Case 2 (Lines \ref{2line:if:2}-\ref{2line:if:3})} If $N-M<M$ (i.e., $N<2M$), $\min\{M,N-M\}=N-M$. Hence greedily  generate $[*]_{M\times (N-M)}$, $[*]_{(N-M)\times M}$,
     and then proceed an $(M,2M-N)$ SDA.
\end{enumerate}
The two cases above are recursively carried out until $N=1$, i.e., $\mathbf{P}=[*]_{1\times 1}$.

\begin{algorithm}[http]
\caption{Greedy SDA Algorithm (G-SDA)}
\label{G-SDA alg}
\begin{algorithmic}[1] 
\REQUIRE  Positive integers $(N,M)$ with $1\leq M\leq N$ and $\gcd(N,M)=1$;
\ENSURE An $(N,M)$ array $\mathbf{P}$ of size $N\times N$;
\STATE \textbf{Procedure} GreedySDA ($N,M$)
\IF {$N=1$}\label{line:case2:1}
\STATE\label{Output:2}
$\mathbf{P}=[*]_{1\times 1}$;
\ELSE
\IF {$N\geq 2M$}\label{line:case3:1}
\STATE \label{line:if:2}$\mathbf{P}'=\textnormal{GreedySDA}\left(N-M,M\right)$;
\STATE\label{P:form}
$\mathbf{P}=
\left[
\begin{array}{c;{1.5pt/1pt}c}
[*]_{M\times M} &  \\ \hdashline[1.5pt/1pt]
& \mathbf{P}'
\end{array}
\right]_{\text{\scriptsize{$N\times N$}}};
$
\ELSE \label{2line:if:1}
\STATE \label{2line:if:2}$\mathbf{P}'=\textnormal{GreedySDA}\big(M,2M-N\big)$;
\STATE \label{2line:if:3}
$\mathbf{P}=
\left[
\begin{array}{c;{1.5pt/1pt}c}
[*]_{M\times (N-M)} & \mathbf{P}'  \\ \hdashline[1.5pt/1pt]
& [*]_{(N-M)\times M}
\end{array}
\right]_{\text{\scriptsize{$N\times N$}}};
$
\ENDIF\label{line:case3:2}
\ENDIF\label{line:case2:2}
\STATE \textbf{end Procedure}
\end{algorithmic}
\end{algorithm}

\begin{Example}\label{exam:RSDA}
The SDA in Example \ref{Example1}  is in fact constructed by the G-SDA algorithm with input parameters $(N=12,M=5)$. The recursive processes are illustrated in  \eqref{exam:steps},
where $\mathbf{P}_i$ $(1\leq i\leq 6)$ is the output array in the $i$-th recursive step with input parameters specified in the brackets to its right.
\newpage
\begin{IEEEeqnarray}{cCcCccC}
\mathbf{P_1}~$(12,5)$&&\mathbf{P}_2~$(7,5)$&&\mathbf{P}_3~$(5,3)$&&\mathbf{P}_4~$(3,1)$\notag \\
    \left[
      \begin{array}{ccccc;{1.5pt/1pt}ccccccc}
        * & * & * & * & * &  &  &  &  &  &  &  \\
        * & * & * & * & * &  &  &  &  &  &  &  \\
        * & * & * & * & * &  &  &  &  &  &  &  \\
        * & * & * & * & * &  &  &  &  &  &  &  \\
        * & * & * & * & * &  &  &  &  &  &  &  \\
        \hdashline[1.5pt/1pt]
         &  &  &  &  & \multicolumn{7}{c}{\multirow{7}{*}{$\mathbf{P}_2$}}  \\
         &  &  &  &  &  \\
         &  &  &  &  &  \\
         &  &  &  &  &  \\
         &  &  &  &  &  \\
         &  &  &  &  &  \\
         &  &  &  &  &  \\
      \end{array}
    \right]_{12\times 12}
    \hspace{-0.9cm}
&\longrightarrow&
\left[
  \begin{array}{cc;{1.5pt/1pt}ccccc}
    * & * & \multicolumn{5}{c}{\multirow{5}{*}{$\mathbf{P}_3$}} \\
    * & * &  \\
    * & * &  \\
    * & * &  \\
    * & * &  \\
    \hdashline[1.5pt/1pt]
     &  & * & * & * & * & *  \\
     &  & * & * & * & * & *  \\
  \end{array}
\right]_{7\times 7}
\hspace{-0.62cm}
&\longrightarrow&
\left[
  \begin{array}{cc;{1.5pt/1pt}ccc}
    * & * & \multicolumn{3}{c}{\multirow{3}{*}{$\mathbf{P}_4$}} \\
    * & * &  \\
    * & * &  \\
    \hdashline[1.5pt/1pt]
     &  & * & * & * \\
     &  & * & * & * \\
  \end{array}
\right]_{5\times 5}
\hspace{-0.62cm}
&\longrightarrow&
\left[
\begin{array}{c;{1.5pt/1pt}c}
* &  \\ \hdashline[1.5pt/1pt]
& \mathbf{P}_5
\end{array}
\right]_{\text{\scriptsize{$3\times 3$}}} \notag\\
&&\mathbf{P}_5~$(2,1)$&&\mathbf{P}_6~$(1,1)$\notag \\
&\longrightarrow&
\left[
\begin{array}{c;{1.5pt/1pt}c}
* &  \\ \hdashline[1.5pt/1pt]
& \mathbf{P}_6
\end{array}
\right]_{\text{\scriptsize{$2\times 2$}}}
\hspace{-0.62cm}
&\longrightarrow&
\left[
  \begin{array}{c}
    *
  \end{array}
\right]_{1\times 1}\label{exam:steps}
\end{IEEEeqnarray}
\end{Example}

\begin{Theorem}\label{thm:O_SDA}
Given any positive integers $N,M$ with $1\leq M\leq N$,
there exists an $(N,M)$ SDA $\mathbf{P}$ with $\eta_{\mathbf{P}}=\eta\big(\frac{N}{\gcd(N,M)},\frac{M}{\gcd(N,M)}\big)$, where $\eta(N,M)$ is recursively defined for any $N,M$ with $1\leq M\leq N$ and $\gcd(N,M)=1$ by
\begin{IEEEeqnarray}{c}\label{sub-packe number}
\eta(N,M)=\left\{\begin{array}{@{}ll}
                    1,&\mathrm{if}~N=1 \\
                  1+\eta\big(N-M,M\big),  &\mathrm{if}~N>1~\mathrm{and}~N\geq2M\\
                  1+\eta(M,2M-N),&\mathrm{if}~N>1~\mathrm{and}~N< 2M
                 \end{array}
\right..
\end{IEEEeqnarray}
\end{Theorem}
\begin{IEEEproof}
For any $N,M$ with $1\leq M\leq N$, the SDA $\mathbf{P}$ in \eqref{SDA:gcd} has $\eta_{\mathbf{P}}=\eta_{\mathbf{P}'}$,
where $\mathbf{P}'$ is an array output by Algorithm \ref{G-SDA alg} with input parameters $\big(\frac{N}{\gcd(N,M)},\frac{M}{\gcd(N,M)}\big)$.
Thus, it is sufficient to prove that the output array of Algorithm \ref{G-SDA alg} is an SDA $\mathbf{P}$ with $\eta_{\mathbf{P}}=\eta(N,M)$ for any $N,M$ with $1\leq M\leq N$ and $\gcd(N,M)=1$.

First of all, for any input parameters $(N,M)$ with $1\leq M\leq N$ and $\gcd(N,M)=1$, we observe two facts from Lines 6 and 9 of Algorithm \ref{G-SDA alg}: $(1)$ During each recursive procedure,
the recursive input parameters $(N,M)$ always maintain the property that $1\leq M\leq N$ and $\gcd(N,M)=1$; $(2)$ $N$ strictly decreases and thus eventually decreases to $1$. Then, the recursive procedure will
terminate at Line 3, i.e., $N=1$. Actually, it is also easy to observe that the recursions of Algorithm \ref{G-SDA alg} happen $\eta(N,M)$ times.


Secondly, it is easy to verify from Lines 7 and 10 that $\mathbf{P}$ satisfies S1 and S2 with parameters $(N,M)$ if and only if $\mathbf{P}'$ satisfies them with parameters $(N-M,M)$ (if $N\geq 2M$) or $(M,2M-N)$ (if $N<2M$). So,
we can easily prove that the output $\mathbf{P}$ is an SDA with $\eta_{\mathbf{P}}$ satisfying \eqref{sub-packe number}   by the induction method.
\end{IEEEproof}

The corollary below follows from  Theorems \ref{relation: SDA and Problem2} and \ref{thm:O_SDA}.

\begin{Corollary}\label{thm:R_SDA}
For any positive integers $N,K,M$  with $M\in[2:N]$,  there exists a capacity-achieving $(\mu=M/N, N,K)$ linear SC-PIR scheme with sub-packetization $\eta\big(\frac{N}{\gcd(N,M)},\frac{M}{\gcd(N,M)}\big)\cdot (M-1)$.
\end{Corollary}

\begin{Remark}\label{comparision:equVSuneq}
Here, we show that the SDA $\mathbf{P}$ constructed in \eqref{SDA:gcd} can further decrease the sub-packetization of capacity-achieving SC-PIR schemes compared to the optimal equal-size sub-packetization $\frac{N(M-1)}{\gcd(N,M)}$ in Theorem \ref{thm:equalsize}.
Since the $(N,M)$ SDA $\mathbf{P}$ has $\frac{N}{\gcd(N,M)}$ columns for any $M\in[2:N]$, $\eta_{\mathbf{P}}=\eta\big(\frac{N}{\gcd(N,M)},\frac{M}{\gcd(N,M)}\big)\leq\frac{N}{\gcd(N,M)}$. Remarkably, it is easy to prove from \eqref{sub-packe number} that the equality (i.e., $\eta_{\mathbf{P}}=\frac{N}{\gcd(N,M)}$) holds if and only if $\frac{M}{\gcd(N,M)}=1$ or $\frac{N-M}{\gcd(N,M)}=1$.
In the other cases, i.e., $\frac{M}{\gcd(N,M)}\neq 1$ and $\frac{N-M}{\gcd(N,M)}\neq 1$, we have $\eta_{\mathbf{P}}<\frac{N}{\gcd(N,M)}$ and thus such SDAs can be used for generating  capacity-achieving SC-PIR schemes with sub-packetization strictly smaller than the optimal equal-size sub-packetization $\frac{N(M-1)}{\gcd(N,M)}$ (e.g.  Example \ref{Example1}).
In addition, when $\frac{M}{\gcd(N,M)}=1$ or $\frac{N-M}{\gcd(N,M)}=1$, it will be shown in Theorem \ref{optimality:G-SDA} that the associated capacity-achieving SC-PIR scheme has the optimal sub-packetization $\frac{N(M-1)}{\gcd(N,M)}$.
\end{Remark}

\subsection{Improved Construction of Storage Design Arrays}\label{Improved-SDA}

Recall that Algorithm \ref{G-SDA alg} always greedily repeats columns in  the current recursive step $l$, which may lead to many
$s_l=1$ in the latter steps and thus results in large $\eta_{\mathbf{P}}$. Particularly, when $N=2M+1$, it generates SDA with $\{s_1=M,s_{2}=\ldots=s_{2+M}=1\}$.
In principle, in order to minimize $\eta_{\mathbf{P}}$ of SDA, it should be better to design repeated columns from a global perspective.
For this case, by decreasing $s_1$ to $M-1$ and increasing some $s_l$ from $1$ to $2$, we are able to present an improved construction to  decrease the sub-packetization.

Before that, it is worthy to point out the following simple property of SDA.
\begin{Lemma}\label{lem:opposite:SDA}
For any $(N,M)$ SDA $\mathbf{P}=[p_{i,j}]_{N\times \frac{N}{\gcd(N,M)}}$, its opposite array $\overline{\mathbf{P}}=[\overline{p}_{i,j}]_{N\times \frac{N}{\gcd(N,M)}}$ defined by
\begin{IEEEeqnarray}{c}\notag
\overline{p}_{i,j}=\left\{\begin{array}{@{}ll}
                           *, &\mathrm{if}~p_{i,j}=\mathrm{NULL}  \\
                           \mathrm{NULL},  &\mathrm{if}~p_{i,j}=*
                          \end{array}
\right.\label{def:oppositeSDA}
\end{IEEEeqnarray}
is an $(N,N-M)$ SDA. Moreover, the number of distinct columns in $\mathbf{P}$ and $\overline{\mathbf{P}}$ are equal, i.e., $\eta_{\mathbf{P}}=\eta_{\overline{\mathbf{P}}}$.
\end{Lemma}

Firstly, given any positive integer $M\geq 2$, we construct an $(N,M)=(2M+1,M)$ SDA  $\mathbf{Q}_M$ of size $(2M+1)\times(2M+1)$ as
\begin{itemize}
  \item If $M$ is even,
  \begin{IEEEeqnarray}{c}\notag
\mathbf{Q}_M=
\left[\begin{array}{c;{1.5pt/1pt}cc;{1.5pt/1pt}c}
\multirow{3}{*}{\text{$[*]_{M\times(M-1)}$}}&  \multicolumn{2}{c;{1.5pt/1pt}}{\multirow{3}{*}{}} & \text{diag$(*)_{2\times 2}$}\\
 & &  & \vdots  \\
& &  & \text{diag}(*)_{2\times 2}  \\ \hdashline[1.5pt/1pt]
\multirow{2}{*}{} & \multicolumn{2}{c;{1.5pt/1pt}}{\text{$[*]_{(\frac{M}{2}+1)\times M}$}} &  \\
\cdashline{2-4}[1.5pt/1pt]
& \multicolumn{1}{c;{1.5pt/1pt}}{\text{$\overline{\text{diag}(*)}_{\frac{M}{2}\times\frac{M}{2}}$}} & \text{$\overline{\text{diag}(*)}_{\frac{M}{2}\times\frac{M}{2}}$} & \text{$[*]_{\frac{M}{2}\times 2}$}
\end{array}
\right]_{\text{\scriptsize{$(2M+1)\times (2M+1)$}}}
\hspace{-3.35cm}
\begin{array}{c}
\multirow{3}{*}{$\left.\begin{array}{c}\\ \\ \\\end{array}\right\}\frac{M}{2}~\text{blocks diag}(*)_{2\times 2}$}
\\ \\ \\ \\ \\
\end{array};
\end{IEEEeqnarray}
 \item If $M$ is odd,
  \begin{IEEEeqnarray}{c}\notag
\mathbf{Q}_M=
\left[\begin{array}{c;{1.5pt/1pt}cc;{1.5pt/1pt}c}
\multirow{2}{*}{\text{$[*]_{M\times(M-1)}$}}&  \multicolumn{2}{c;{1.5pt/1pt}}{\multirow{2}{*}{}} & \text{diag$(*)_{3\times 3}$}\\
 & &  & \text{$\vdots$}  \\
\cdashline{1-3}[1.5pt/1pt]
\multirow{2}{*}{} & \multicolumn{2}{c;{1.5pt/1pt}}{\text{$[*]_{\frac{M+3}{2}\times (M-1)}$}} & \text{diag}(*)_{3\times 3} \\
\cdashline{2-4}[1.5pt/1pt]
& \multicolumn{1}{c;{1.5pt/1pt}}{\text{$\overline{\text{diag}(*)}_{\frac{M-1}{2}\times\frac{M-1}{2}}$}} & \text{$\overline{\text{diag}(*)}_{\frac{M-1}{2}\times\frac{M-1}{2}}$} & \text{$[*]_{\frac{M-1}{2}\times 3}$}
\end{array}
\right]_{\text{\scriptsize{$(2M+1)\times (2M+1)$}}}
\hspace{-3.35cm}
\begin{array}{c}
\multirow{2}{*}{$\left.\begin{array}{c}\\ \\ \\\end{array}\right\}\frac{M+1}{2}~\text{blocks diag}(*)_{3\times 3}$}\\
\\
\\
\\
\end{array},
\end{IEEEeqnarray}
\end{itemize}
where $\mathrm{diag}(*)_{n\times n}$ denotes an $n\times n$ array with the entries in  diagonal being ``$*$"s and the rest entries being ``$\mathrm{NULL}$"s.
It is easily checked that $\mathbf{Q}_M$ is a $(2M+1,M)$ SDA with $\eta_{\mathbf{Q}_M}=\big\lceil\frac{M}{2}\big\rceil+3$.

\begin{Example}
When $M=4$ and $M=5$, $\mathbf{Q}_4$ and $\mathbf{Q}_5$ are the following forms, respectively.
\begin{IEEEeqnarray}{c}\notag
\mathbf{Q}_4=\left[
  \begin{array}{ccc;{1.5pt/1pt}cccc;{1.5pt/1pt}cc}
   * & * & * &  &  &  &  & * &   \\
   * & * & * &  &  &  &  &  & *  \\
   * & * & * &  &  &  &  & * &   \\
   * & * & * &  &  &  &  &  & * \\ \cdashline{1-9}[1.5pt/1pt]
    &  &  & * & * & * & * &  &  \\
    &  &  & * & * & * & * &  &   \\
    &  &  & * & * & * & * &  &   \\ \cdashline{4-9}[1.5pt/1pt]
    &  &  &  & \multicolumn{1}{c;{1.5pt/1pt}}{*} &  & * & * & *  \\
    &  &  & * & \multicolumn{1}{c;{1.5pt/1pt}}{} & * &  & * & * \\
  \end{array}
\right]_{9\times 9},\quad
\mathbf{Q}_5=
\left[
  \begin{array}{cccc;{1.5pt/1pt}cccc;{1.5pt/1pt}ccc}
  * & * & * & * &  &  &  &  & * &  &  \\
  * & * & * & * &  &  &  &  &  & * &  \\
  * & * & * & * &  &  &  &  &  &  & * \\
  * & * & * & * &  &  &  &  & * &  &  \\
  * & * & * & * &  &  &  &  &  & * &  \\ \cdashline{1-8}[1.5pt/1pt]
   &  &  &  & * & * & * & * &  &  & * \\
   &  &  &  & * & * & * & * & * &  &  \\
   &  &  &  & * & * & * & * &  & * &  \\
   &  &  &  & * & * & * & * &  &  & * \\ \cdashline{5-11}[1.5pt/1pt]
   &  &  &  &  & \multicolumn{1}{c;{1.5pt/1pt}}{*} &  & * & * & * & * \\
   &  &  &  & * & \multicolumn{1}{c;{1.5pt/1pt}}{} & * &  & * & * & * \\
  \end{array}
\right]_{11\times 11}.
\end{IEEEeqnarray}
Obviously, $\eta_{\mathbf{Q}_4}=5$ with $\{s_1=3,s_2=2,s_3=2,s_4=1,s_5=1\}$ and $\eta_{\mathbf{Q}_5}=6$ with $\{s_1=4,s_2=2,s_3=2,s_4=1,s_5=1,s_6=1\}$.
Compared to $\mathbf{P}$ with $\{s_1=4,s_2=1,s_3=1,s_4=1,s_5=1,s_6=1\}$ (resp. $\mathbf{P}'$ with $\{s_1=5,s_2=1,s_3=1,s_4=1,s_5=1,s_6=1,s_7=1\}$) in Example \ref{SDA_def},
the distribution of repeated columns $s_l$ is more flexible than the greedy algorithm, which leads to a smaller number of distinct columns.
\end{Example}

Notice that $\mathbf{Q}_{M-1}$ is a $(2M-1,M-1)$ SDA with $\eta_{\mathbf{Q}_{M-1}}=\big\lfloor\frac{M}{2}\big\rfloor+3$ for any $M\geq 3$. By Lemma \ref{lem:opposite:SDA}, we can obtain a $(2M-1,M)$ SDA $\overline{\mathbf{Q}}_{M-1}$ with $\eta_{\overline{\mathbf{Q}}_{M-1}}=\eta_{\mathbf{Q}_{M-1}}=\big\lfloor\frac{M}{2}\big\rfloor+3$ for any $M\geq 3$. Next, based on $\mathbf{Q}_{M}$ and $\overline{\mathbf{Q}}_{M-1}$, for any given positive integers $M,d$ such that $M\geq 3,d\geq 2$, we can construct a class of $(N,M)$ SDA with $N=dM\pm1$ as:
\begin{IEEEeqnarray}{c}
\mathbf{P}=\substack{\overbrace{~~~~~~~~~~~~~~~~~~~~~~~~~~~~~~~~~~~~~~~~~}^{\text{\normalsize{$d-2$}}}~~~~~~~~~~~~\\
\left[\begin{array}{cccc;{1.5pt/1pt}c}
                          [*]_{M\times M} &  &  & &\multirow{4}{*}{}  \\
                           \text{} &  [*]_{M\times M}&  &  &   \\
                            &  & \ddots &  & \\
                           \text{}& \text{} &  &  [*]_{M\times M} &  \\  \hdashline[1.5pt/1pt]
                            \multicolumn{4}{c;{1.5pt/1pt}}{\text{}}  & \mathbf{P}'
                         \end{array}
\right]_{\text{\scriptsize{$N\times N$}}}},\label{improved:SDA}
\end{IEEEeqnarray}
where
\begin{IEEEeqnarray}{c}\notag
\mathbf{P}'=\left\{\begin{array}{@{}ll}
                      \mathbf{Q}_M, &\mathrm{if}~N=dM+1 \\
                      \overline{\mathbf{Q}}_{M-1}, &\mathrm{if}~N=dM-1
                    \end{array}
\right..
\end{IEEEeqnarray}

The following result is straightforward.
\begin{Theorem}\label{thm:improved:SDA1}
Given any positive integer $N=d M\pm 1$ for  integers $M\in [3:N]$ and $d\geq 2$, there is  an $(N,M)$ SDA $\mathbf{P}$
with
\begin{IEEEeqnarray}{rCl}\notag
\eta_\mathbf{P}&=&\left\{\begin{array}{@{}ll}
            d+\big\lceil\frac{M}{2}\big\rceil+1, &\mathrm{if}~N=dM+1  \\
           d+\big\lfloor\frac{M}{2}\big\rfloor+1,  &\mathrm{if}~N=dM-1
          \end{array}
\right..\label{eqn:eta:Pimprove}
\end{IEEEeqnarray}
\end{Theorem}

\begin{Remark}\label{remark:improve}
Notice that $\gcd(N,M)=1$ when $N=dM\pm 1$ for any integer $d\geq 2$. Let $(N,M)$ be the input parameters of the G-SDA algorithm, then the number of distinct columns of the output SDA is equal to $\eta(N,M)$ in  \eqref{sub-packe number} at $N=dM\pm 1$, i.e.,
\begin{IEEEeqnarray}{c} \notag
\eta(N,M)=\left\{\begin{array}{@{}ll}
                    d+M,&\mathrm{if}~N=dM+1\\
                  d+M-1,  &\mathrm{if}~N=dM-1
                 \end{array}
\right..
\end{IEEEeqnarray}
Thus, the SDA presented in \eqref{improved:SDA} decreases the number of distinct columns and further reduces the required sub-packetization of capacity-achieving SC-PIR scheme.
\end{Remark}

Finally,   we obtain the following corollary from Theorems \ref{relation: SDA and Problem2} and \ref{thm:improved:SDA1}.

\begin{Corollary}\label{thm:improved:SDA}
Given any positive integers $N,K,M$  with  $M\in[3:N]$ and $N=d M\pm 1$ for some integer $d\geq 2$, there exists a capacity-achieving $(\mu=M/N,N,K)$ linear SC-PIR scheme with sub-packetization $\big(d+\big\lceil\frac{M}{2}\big\rceil+1\big)\cdot(M-1)$ if $N=dM+1$ or $\big(d+\big\lfloor \frac{M}{2}\big\rfloor+1\big)\cdot(M-1)$ if $N=dM-1$.
\end{Corollary}

In general, it is difficult to  extend the above  improvement  to all the parameters $N$ and $M\in[1:N]$, since the modification of $s_l$ becomes very complicated and sometimes even unsolvable under the SDA constraints S1 and S2 if there are many different values of $s_l$.

\subsection{Optimality on Sub-packetization of Greedy SDA}\label{lower bound}

In this subsection, firstly we establish a lower bound  on the optimal value of Problem 1. Next,
we discuss the optimality on sub-packetization of the SC-PIR scheme associated to the proposed SDA with respect to this bound.

\begin{Lemma}\label{optimality:G-SDA-0}
For any positive integers $N,M$ with $2\leq M<N$, the optimal value of Problem 1 satisfies
\begin{eqnarray}\label{lower bound:Problem 1}
 \eta^{*}\geq\max\left\{\left\lceil\frac{N}{M}\right\rceil,\left\lceil\frac{N}{N-M}\right\rceil\right\}.
 \end{eqnarray}
\end{Lemma}

\begin{IEEEproof}
Suppose that $\{\alpha_{\mathcal{S}}^*\}_{\mathcal{S}\subseteq[1:N],|\mathcal{S}|=M}$ is the optimal solution to Problem 1. Let $\{\mathcal{S}_i\}_{i=1}^{\eta^*}$ be the indices of non-zero elements in the solution,  i.e., $\alpha_{\mathcal{S}}^{*}>0$ if and only if $\mathcal{S}\in\{\mathcal{S}_{i}\}_{i=1}^{\eta^*}$.
By \eqref{sum:alpha:T}, the optimal solution satisfying \eqref{optimize:condition1} and \eqref{optimize:condition2} implies the following constraint
\begin{IEEEeqnarray}{c}\label{sum:constraint}
\sum\limits_{\substack{\mathcal{S}\subseteq[1:N]\\|\mathcal{S}|=M}}\alpha_{\mathcal{S}}^{*}=\sum\limits_{i\in[1:\eta^{*}]}\alpha_{\mathcal{S}_i}^{*}=1.
\end{IEEEeqnarray}
Hence, there must be an index $j\in[1:\eta^{*}]$ such that
\begin{IEEEeqnarray}{c}
\alpha_{\mathcal{S}_{j}}^{*}\geq\frac{1}{\eta^{*}}.\label{eqn:alp:Sj}
\end{IEEEeqnarray}

Assume that $\eta^{*}<\big\lceil\frac{N}{M}\big\rceil$. Then, $\frac{N}{M}-\eta^{*}>0$ since $\eta^{*}$ is an integer. Accordingly, there must exist a positive number $c=\frac{N}{M}-\eta^{*}>0$. By \eqref{eqn:alp:Sj},
\begin{IEEEeqnarray}{rCl}
\alpha_{\mathcal{S}_{j}}^{*}
&\geq&\frac{1}{\frac{N}{M}-c} \notag\\
&>&\frac{M}{N} \notag\\
&=&\mu. \notag
\end{IEEEeqnarray}
Thus, for any $n\in\mathcal{S}_j$,
\begin{IEEEeqnarray}{rCl}
\sum\limits_{\substack{\mathcal{S}\subseteq[1:N]\\|\mathcal{S}|=M,n\in\mathcal{S}}}\alpha_{\mathcal{S}}^{*} 
\geq \alpha_{\mathcal{S}_j}^{*}
>\mu, \label{proof:improved}
\end{IEEEeqnarray}
which  contradicts  \eqref{optimize:condition1}. Thus, $\eta^*\geq\big\lceil\frac{N}{M}\big\rceil$.

Assume that $\eta^{*}< \big\lceil\frac{N}{N-M}\big\rceil$. Similarly, we have $\eta^{*}=\frac{N}{N-M}-c$ for some $c>0$. Then,
\begin{IEEEeqnarray}{rCl}
\alpha_{\mathcal{S}_{j}}^{*}
&\geq&\frac{1}{\frac{N}{N-M}-c} \notag\\
&>&\frac{N-M}{N} \notag\\
&=&1-\mu.\notag\label{second:max}
\end{IEEEeqnarray}
Since $M<N$,    there exists $n\in[1:N]\backslash\mathcal{S}_j$ such that
\begin{IEEEeqnarray}{rCl}
\sum\limits_{\substack{\mathcal{S}\subseteq[1:N]\\|\mathcal{S}|=M,n\in\mathcal{S}}}\alpha_{\mathcal{S}}^{*}
&=&\sum\limits_{\substack{\mathcal{S}\subseteq[1:N],|\mathcal{S}|=M\\ \mathcal{S}\neq\mathcal{S}_j,n\in\mathcal{S}}}\alpha_{\mathcal{S}}^{*} \notag\\
&\overset{(a)}{\leq} &1-\alpha_{\mathcal{S}_j}^{*} \notag\\
&{<}&\mu, \notag
\end{IEEEeqnarray}
where  $(a)$ is due to \eqref{sum:constraint},  a contradiction to  \eqref{optimize:condition1} again. That is, $\eta^*\geq\big\lceil\frac{N}{N-M}\big\rceil$.
\end{IEEEproof}

Recall from Theorem \ref{thm:O_SDA} and Corollary \ref{thm:R_SDA} that the capacity-achieving linear SC-PIR scheme associated to the SDA in \eqref{SDA:gcd} has sub-packetization
 \begin{IEEEeqnarray}{c}\label{SDA:sub}
 F_{\text{G-SDA}}^{(N,M)}\triangleq\eta\left(\frac{N}{\gcd(N,M)},\frac{M}{\gcd(N,M)}\right)\cdot(M-1).
 \end{IEEEeqnarray}

In general, it is difficult to directly compare the sub-packetization $F_{\text{G-SDA}}^{(N,M)}$ and the optimal value $F^{*}=\eta^{*}\cdot (M-1)$  in
Theorem \ref{arbitra:part2} since  $\eta\big(\frac{N}{\gcd(N,M)},\frac{M}{\gcd(N,M)}\big)$ is defined in \eqref{sub-packe number} as a \emph{recursive} form.
In the following theorem, we obtain a multiplicative gap by  relaxing $\eta\big(\frac{N}{\gcd(N,M)},\frac{M}{\gcd(N,M)}\big)$ to an upper bound and $\eta^{*}$ to a lower bound.
In addition, this theorem also characterize two special optimal cases.

\begin{Theorem}\label{optimality:G-SDA}
Given any $(\mu,N,K)$ SC-PIR system with $M=\mu N\in[2:N]$, the sub-packetization $F_{\textnormal{G-SDA}}^{(N,M)}$ is within a multiplicative gap $\frac{\min\{M,N-M\}}{\gcd(N,M)}$ of the optimal sub-packetization of capacity-achieving linear SC-PIR schemes. In particular, in the cases $\min\{M,N-M\}|N$ or $M=N$, the sub-packetization $F_{\textnormal{G-SDA}}^{(N,M)}=\frac{N(M-1)}{\gcd(N,M)}$ is optimal.
\end{Theorem}

\begin{IEEEproof}
From Theorem \ref{arbitra:part2}, the optimal sub-packetization is $F^{*}=\eta^{*}\cdot (M-1)$.
In the case $M=N$, since $\eta^*\geq 1$, $F_{\textnormal{G-SDA}}^{(N,M)}=M-1=F^*$ is optimal.
When $M\in[2: N-1]$, 
\begin{IEEEeqnarray}{rCl}
1&\leq&\frac{F_{\textnormal{G-SDA}}^{(N,M)}}{F^*} \notag\\
&\overset{(a)}{=}&\frac{\eta\big(\frac{N}{\gcd(N,M)},\frac{M}{\gcd(N,M)}\big)}{\eta^{*}}\\
&\overset{(b)}{\leq}&\frac{\frac{N}{\gcd(N,M)}}{\max\big\{\big\lceil\frac{N}{M}\big\rceil,\big\lceil\frac{N}{N-M}\big\rceil\big\}} \notag\\
&\leq&\frac{\min\{M,N-M\}}{\gcd(N,M)},\label{eqn:optimality}
\end{IEEEeqnarray}
where $(a)$ follows by \eqref{SDA:sub};
$(b)$ is due to the lower bound in \eqref{lower bound:Problem 1} and the upper bound $\eta\big(\frac{N}{\gcd(N,M)},\frac{M}{\gcd(N,M)}\big)\leq\frac{N}{\gcd(N,M)}$ by Remark \ref{comparision:equVSuneq}.

%
%
In particular, in the case $\min\{M,N-M\}|N$, we have $\gcd(N,M)=\min\{M,N-M\}$. Thus, $F_{\textnormal{G-SDA}}^{(N,M)}=F^*$ by \eqref{eqn:optimality}, i.e., $F_{\textnormal{G-SDA}}^{(N,M)}=\frac{N}{\gcd(N,M)}\cdot(M-1)$ is the optimal sub-packetization in this case.
\end{IEEEproof}

\begin{Remark}\label{remark 4}
The SC-PIR problem straightly degrades to the problem of replicated PIR by setting $M=N$. 
 Obviously, when $M=N$, our SC-PIR scheme associated to the SDA in \eqref{SDA:gcd} achieves the optimal sub-packetization $F^{*}=M-1=N-1$, which is the same as that in \cite{Tian and Sun}.
\end{Remark}

\section{Conclusion}\label{conclusion}

In this paper, we investigated the sub-packetization of uncoded Storage Constrained PIR (SC-PIR).
We first characterized the optimal sub-packetization of capacity-achieving SC-PIR schemes by an optimization problem, 
which is hard to solve due to the involved  non-continuous indicator functions.
We introduced Storage Design Array (SDA) to construct practical operational SC-PIR schemes with low sub-packetization.
It turns out that, for the SC-PIR system with $N$ servers and a total normalized storage capacity $M\in\{2,\ldots,N\}$, the equal-size sub-packetization $\frac{N(M-1)}{\gcd(N,M)}$ is optimal among all  capacity-achieving linear SC-PIR schemes characterized by Woolsey \emph{et al.}.
Finally, by allowing unequal-size packets, two constructions of SDAs were proposed to further decrease the sub-packetization.
The resultant sub-packetizations were shown to be optimal in the cases $\min\{M,N-M\}|N$ or $M=N$, and within a multiplicative gap of $\frac{\min\{M,N-M\}}{\gcd(N,M)}$ compared to a lower bound on the optimal sub-packetization otherwise.

\section*{Appendix}

In this section, we provide the proof of the five necessary conditions P1--P5 for any capacity-achieving linear SC-PIR scheme. To this end,
we first refine the converse proof of SC-PIR capacity given in \cite{Attia SC-PIR}, and accordingly, we have placed an emphasis on the necessary properties of capacity-achieving SC-PIR schemes by constraining the inequalities in the refined proof to be held with equalities.
Further, the obtained properties are specialized to linear SC-PIR schemes to complete the proof.
Actually, the similar approach was used in \cite{Tian and Sun} and \cite{Zhu} for the setups of replicated PIR and MDS-coded PIR, respectively.

We start by proving two useful lemmas.
\subsection{Preliminary Lemmas}
\begin{Lemma}
For any $i\in[1:N],\mathcal{K}\subseteq[1:K]$, and $\mathcal{N}\subseteq[1:N]$,
\begin{IEEEeqnarray}{c}\label{Appen:query}
H(A_{i}^{[\theta]}|{W}_{\mathcal{K}},{Z}_{\mathcal{N}},Q_{i}^{[\theta]})=H(A_{i}^{[\theta]}|{W}_{\mathcal{K}},{Z}_{\mathcal{N}},{Q}_{1:N}^{[\theta]}),\quad\forall \,\theta\in[1:K].
\end{IEEEeqnarray}
\end{Lemma}
\begin{IEEEproof}
Notice that
\begin{IEEEeqnarray}{rCl}
0&\leq&H(A_{i}^{[\theta]}|{W}_{\mathcal{K}},{Z}_{\mathcal{N}},Q_{i}^{[\theta]})-H(A_{i}^{[\theta]}|{W}_{\mathcal{K}},{Z}_{\mathcal{N}},{Q}_{1:N}^{[\theta]}) \notag \\
&=&I(A_{i}^{[\theta]};{Q}_{[1:N]\backslash\{i\}}^{[\theta]}|{W}_{\mathcal{K}},{Z}_{\mathcal{N}},Q_{i}^{[\theta]}) \notag \\
&\leq&I(A_{i}^{[\theta]},{W}_{[1:K]\backslash\mathcal{K}};{Q}_{[1:N]\backslash\{i\}}^{[\theta]}|{W}_{\mathcal{K}},{Z}_{\mathcal{N}},Q_{i}^{[\theta]}) \notag \\
&=&I({W}_{[1:K]\backslash\mathcal{K}};{Q}_{[1:N]\backslash\{i\}}^{[\theta]}|{W}_{\mathcal{K}},{Z}_{\mathcal{N}},Q_{i}^{[\theta]})
+I(A_{i}^{[\theta]};{Q}_{[1:N]\backslash\{i\}}^{[\theta]}|{W}_{1:K},{Z}_{\mathcal{N}},Q_{i}^{[\theta]}) \notag \\
&\overset{(a)}{=}&I({W}_{[1:K]\backslash\mathcal{K}};{Q}_{[1:N]\backslash\{i\}}^{[\theta]}|{W}_{\mathcal{K}},{Z}_{\mathcal{N}},Q_{i}^{[\theta]}) \notag \\
&\leq&I(W_{1:K},{Z}_{\mathcal{N}};Q_{[1:N]\backslash\{i\}}^{[\theta]}|Q_i^{[\theta]})\notag\\
&\leq&I(W_{1:K},{Z}_{\mathcal{N}};Q_{1:N}^{[\theta]})\notag\\
&\overset{(b)}{=}&I(W_{1:K};Q_{1:N}^{[\theta]})\notag\\
&\overset{(c)}{=}&0, \notag
\end{IEEEeqnarray}
where $(a)$ holds because $A_{i}^{[\theta]}$ is a function of ${W}_{1:K}$ and $Q_{i}^{[\theta]}$ by \eqref{model:answers}; 
$(b)$ follows from the fact that ${Z}_{\mathcal{N}}$ is a function of the files $W_{1:K}$ by \eqref{file:same_storage}; and $(c)$ follows from the  independence of files and queries by \eqref{model:query inden}.
\end{IEEEproof}

\begin{Lemma}\label{appendix privacy}
For any $i\in[1:N],\mathcal{K}\subseteq[1:K]$, and $\mathcal{N}\subseteq[1:N]$,
\begin{IEEEeqnarray}{c}\label{Appen:priva proof}
H(A_{i}^{[\theta]}|{W}_{\mathcal{K}},{Z}_{\mathcal{N}},Q_{i}^{[\theta]})=H(A_{i}^{[\theta']}|{W}_{\mathcal{K}},{Z}_{\mathcal{N}},{Q}_{i}^{[\theta']}),\quad\forall\, \theta,\theta'\in[1:K].
\end{IEEEeqnarray}
\end{Lemma}
\begin{IEEEproof}
For any $\theta\in[1:K]$,
\begin{IEEEeqnarray}{rCl}
0&\leq& I(Q_{i}^{[\theta]},A_{i}^{[\theta]},{W}_{\mathcal{K}},{Z}_{\mathcal{N}};\theta)\notag\\
&\leq&I(Q_{i}^{[\theta]},A_{i}^{[\theta]},{W}_{1:K},{Z}_{\mathcal{N}};\theta) \notag\\
&\overset{(a)}{=}&I(Q_{i}^{[\theta]},{W}_{1:K};\theta)\notag\\
&=&I(Q_{i}^{[\theta]};\theta)+I({W}_{1:K};\theta|Q_i^{[\theta]})\notag\\
&\overset{(b)}{=}&I(Q_i^{[\theta]}; \theta)\notag\\
&\leq &I(Q_i^{[\theta]},A_i^{[\theta]},Z_i;\theta)\notag\\
&\overset{(c)}{=}&0,\notag
\end{IEEEeqnarray}
where $(a)$ holds because $A_i^{[\theta]}$ and ${Z}_\mathcal{N}$ are determined by $Q_i^{[\theta]}$ and $W_{1:K}$, $(b)$ holds because the files are independent of the desired file index and the query, and $(c)$ follows from  the privacy constraint \eqref{Infor:priva cons}.

Notice that, $I(Q_{i}^{[\theta]},A_{i}^{[\theta]},{W}_{\mathcal{K}},{Z}_{\mathcal{N}};\theta)=0$
indicates that $(Q_{i}^{[\theta]},A_{i}^{[\theta]},{W}_{\mathcal{K}},{Z}_{\mathcal{N}})$ and $\theta$ are independent of each other. Therefore, \eqref{Appen:priva proof} holds.
\end{IEEEproof}

Next, we present some properties of any SC-PIR scheme. 
\subsection{Properties of SC-PIR Schemes}
For $n\in[0:N-1]$ and $k\in[0:K-1]$, define
\begin{IEEEeqnarray}{c}\label{defi:Tnk}
T(n,k)\triangleq \frac{1}{NK\tbinom{K-1}{k}\tbinom{N-1}{n}}\sum\limits_{\substack{\mathcal{K}\subseteq[1:K]\\|\mathcal{K}|=k}} \sum\limits_{\substack{\mathcal{N}\subseteq[1:N]\\|\mathcal{N}|=n}}
\sum\limits_{\theta\in[1:K]\backslash\mathcal{K}} \sum\limits_{i\in[1:N]\backslash\mathcal{N}} H(A_{i}^{[\theta]}|{W}_{\mathcal{K}},{Z}_{\mathcal{N}},Q_{i}^{[\theta]})
\end{IEEEeqnarray}
and
\begin{IEEEeqnarray}{c}\label{condi:intial2}
T(n,K)\triangleq0,\quad\forall\, n\in[0:N-1].
\end{IEEEeqnarray}
In addition,
\begin{IEEEeqnarray}{c}\label{Defi:lambda}
\lambda_{n}\triangleq\frac{1}{K\tbinom{N}{n}}\sum\limits_{\substack{\mathcal{N}\subseteq[1:N]\\|\mathcal{N}|=n}}\sum\limits_{\theta\in[1:K]}H(W_{\theta}|{Z}_{\mathcal{N}}),\quad\forall\, n\in[0:N-1].
\end{IEEEeqnarray}

\begin{Lemma}\label{recr:lemma}
For each $n\in[0:N-1]$ and $k\in[0:K-1]$, any SC-PIR scheme must satisfy the following inductive relationship.
\begin{IEEEeqnarray}{c}\label{Lemma:rescu}
T(n,k)\geq\frac{1}{N-n}\left[ \sum\limits_{n'=n}^{N-1}T(n',k+1)+\lambda_{n} \right].
\end{IEEEeqnarray}
Moreover, to establish the equality in \eqref{Lemma:rescu}, 
 for every realization of queries $\widetilde{{Q}}_{1:N}^{[\theta]}$ with positive probability,
\begin{itemize}
  \item For any $\mathcal{K}\subseteq[1:K]$ of size $k$, $\mathcal{N}\subseteq[1:N]$ of size $n$, and $\theta\in[1:K]\backslash\mathcal{K}$,
    \begin{IEEEeqnarray}{c}\label{C11}
    \sum\limits_{i\in[1:N]\backslash\mathcal{N}} H(A_{i}^{[\theta]}|{W}_{\mathcal{K}},{Z}_{\mathcal{N}},{Q}_{1:N}^{[\theta]}=\widetilde{{Q}}_{1:N}^{[\theta]})= H({A}_{[1:N]\backslash\mathcal{N}}^{[\theta]}|{W}_{\mathcal{K}},{Z}_{\mathcal{N}},{Q}_{1:N}^{[\theta]}=\widetilde{{Q}}_{1:N}^{[\theta]});
    \end{IEEEeqnarray}
    \item For any $\mathcal{K}\subseteq[1:K], \theta\in\mathcal{K},\mathcal{N}_1\subseteq[1:N],\mathcal{N}_2\subseteq[1:N]\backslash\mathcal{N}_1$ and $i\in[1:N]\backslash(\mathcal{N}_1\cup\mathcal{N}_2)$ such that $|\mathcal{K}|=k+1$, $|\mathcal{N}_1|=n$, $|\mathcal{N}_2|=n'-n$ with $n'\in[n+1:N-1]$,
    \begin{IEEEeqnarray}{c}\label{C22}
    I(A_{i}^{[\theta]};{Z}_{\mathcal{N}_2}|{W}_{\mathcal{K}},{Z}_{\mathcal{N}_1},{A}_{\mathcal{N}_2}^{[\theta]},{Q}_{1:N}^{[\theta]}=\widetilde{{Q}}_{1:N}^{[\theta]})=0.
    \end{IEEEeqnarray}
\end{itemize}
\end{Lemma}
\begin{IEEEproof}
For any $n\in[0:N-1]$ and $k\in[0:K-1]$, $T(n,k)$ in \eqref{defi:Tnk} can be lower bounded as
\begin{IEEEeqnarray}{rCl}
T(n,k)&=& \frac{1}{NK\tbinom{K-1}{k}\tbinom{N-1}{n}}\sum\limits_{\substack{\mathcal{K}\subseteq[1:K]\\|\mathcal{K}|=k}} \sum\limits_{\substack{\mathcal{N}\subseteq[1:N]\\|\mathcal{N}|=n}}
\sum\limits_{\theta\in[1:K]\backslash\mathcal{K}} \sum\limits_{i\in[1:N]\backslash\mathcal{N}} H(A_{i}^{[\theta]}|{W}_{\mathcal{K}},{Z}_{\mathcal{N}},Q_{i}^{[\theta]}) \notag \\
&\overset{(a)}{=}& \frac{1}{NK\tbinom{K-1}{k}\tbinom{N-1}{n}}\sum\limits_{\substack{\mathcal{K}\subseteq[1:K]\\|\mathcal{K}|=k}} \sum\limits_{\substack{\mathcal{N}\subseteq[1:N]\\|\mathcal{N}|=n}}
\sum\limits_{\theta\in[1:K]\backslash\mathcal{K}} \sum\limits_{i\in[1:N]\backslash\mathcal{N}} H(A_{i}^{[\theta]}|{W}_{\mathcal{K}},{Z}_{\mathcal{N}},{Q}_{1:N}^{[\theta]}) \notag \\
&\overset{(b)}{\geq}& \frac{1}{NK\tbinom{K-1}{k}\tbinom{N-1}{n}}\sum\limits_{\substack{\mathcal{K}\subseteq[1:K]\\|\mathcal{K}|=k}} \sum\limits_{\substack{\mathcal{N}\subseteq[1:N]\\|\mathcal{N}|=n}}
\sum\limits_{\theta\in[1:K]\backslash\mathcal{K}} H({A}_{[1:N]\backslash\mathcal{N}}^{[\theta]}|{W}_{\mathcal{K}},{Z}_{\mathcal{N}},{Q}_{1:N}^{[\theta]}) \label{pro:Lemma1} \\
&\overset{(c)}{=}& \frac{1}{NK\tbinom{K-1}{k}\tbinom{N-1}{n}}\sum\limits_{\substack{\mathcal{K}\subseteq[1:K]\\|\mathcal{K}|=k}} \sum\limits_{\substack{\mathcal{N}\subseteq[1:N]\\|\mathcal{N}|=n}}
\sum\limits_{\theta\in[1:K]\backslash\mathcal{K}}\left[H({A}_{[1:N]\backslash\mathcal{N}}^{[\theta]}|{W}_{\mathcal{K}},{Z}_{\mathcal{N}},{Q}_{1:N}^{[\theta]})  +H(W_{\theta}|{A}_{[1:N]\backslash\mathcal{N}}^{[\theta]},{W}_{\mathcal{K}},{Z}_{\mathcal{N}},{Q}_{1:N}^{[\theta]})\right] \notag \\
&=& \frac{1}{NK\tbinom{K-1}{k}\tbinom{N-1}{n}}\sum\limits_{\substack{\mathcal{K}\subseteq[1:K]\\|\mathcal{K}|=k}} \sum\limits_{\substack{\mathcal{N}\subseteq[1:N]\\|\mathcal{N}|=n}}
\sum\limits_{\theta\in[1:K]\backslash\mathcal{K}} H({A}_{[1:N]\backslash\mathcal{N}}^{[\theta]},W_{\theta}|{W}_{\mathcal{K}},{Z}_{\mathcal{N}},{Q}_{1:N}^{[\theta]}) \notag \\
&\overset{(d)}{=}& \frac{1}{NK\tbinom{K-1}{k}\tbinom{N-1}{n}}\sum\limits_{\substack{\mathcal{K}\subseteq[1:K]\\|\mathcal{K}|=k}} \sum\limits_{\substack{\mathcal{N}\subseteq[1:N]\\|\mathcal{N}|=n}}
\sum\limits_{\theta\in[1:K]\backslash\mathcal{K}} H({A}_{[1:N]\backslash\mathcal{N}}^{[\theta]}|{W}_{(\mathcal{K}\cup\{\theta\})},{Z}_{\mathcal{N}},{Q}_{1:N}^{[\theta]}) \notag \\
&&+\frac{1}{NK\tbinom{K-1}{k}\tbinom{N-1}{n}}\sum\limits_{\substack{\mathcal{K}\subseteq[1:K]\\|\mathcal{K}|=k}} \sum\limits_{\substack{\mathcal{N}\subseteq[1:N]\\|\mathcal{N}|=n}}
\sum\limits_{\theta\in[1:K]\backslash\mathcal{K}} H(W_{\theta}|{Z}_{\mathcal{N}}) \notag \\
&\overset{(e)}{=}& \underbrace{\frac{1}{NK\tbinom{K-1}{k}\tbinom{N-1}{n}}\sum\limits_{\substack{\mathcal{K}\subseteq[1:K]\\|\mathcal{K}|=k+1}} \sum\limits_{\substack{\mathcal{N}\subseteq[1:N]\\|\mathcal{N}|=n}}
\sum\limits_{\theta\in\mathcal{K}} H({A}_{[1:N]\backslash\mathcal{N}}^{[\theta]}|{W}_{\mathcal{K}},{Z}_{\mathcal{N}},{Q}_{1:N}^{[\theta]})}_{\triangleq \widetilde{T}(n,k)} \notag \\
&&+\frac{1}{N-n}\cdot\frac{1}{K\tbinom{K-1}{k}\tbinom{N}{n}}\sum\limits_{\substack{\mathcal{K}\subseteq[1:K]\\|\mathcal{K}|=k}} \sum\limits_{\substack{\mathcal{N}\subseteq[1:N]\\|\mathcal{N}|=n}}
\sum\limits_{\theta\in[1:K]\backslash\mathcal{K}} H(W_{\theta}|{Z}_{\mathcal{N}}) \notag \\
&\overset{(f)}{=}&\widetilde{T}(n,k)+\frac{1}{N-n}\cdot\frac{1}{K\tbinom{K-1}{k}\tbinom{N}{n}}\sum\limits_{\substack{\mathcal{N}\subseteq[1:N]\\|\mathcal{N}|=n}}
\sum\limits_{\theta\in[1:K]}\sum\limits_{\substack{\mathcal{K}\subseteq[1:K]\backslash\{\theta\}\\|\mathcal{K}|=k}} H(W_{\theta}|{Z}_{\mathcal{N}}) \notag \\
&\overset{(g)}{=}&\widetilde{T}(n,k)+\frac{1}{N-n}\cdot\frac{1}{K\tbinom{N}{n}}\sum\limits_{\substack{\mathcal{N}\subseteq[1:N]\\|\mathcal{N}|=n}}\sum\limits_{\theta\in[1:K]}H(W_{\theta}|{Z}_{\mathcal{N}}) \notag\\
&\overset{(h)}{=}&\widetilde{T}(n,k)+\frac{1}{N-n}\cdot\lambda_{n}, \label{appen:lower bound1}
\end{IEEEeqnarray}
where $(a)$ follows by \eqref{Appen:query};
$(b)$ follows from the property of independence bound on entropy;
$(c)$ follows from  \eqref{model:answers} in which ${A}_{\mathcal{N}}^{[\theta]}$ are a determined function of ${Z}_{\mathcal{N}}$ and ${Q}_{\mathcal{N}}^{[\theta]}$, thus with  $({A}_{[1:N]\backslash\mathcal{N}}^{[\theta]},{Z}_{\mathcal{N}},{Q}_{1:N}^{[\theta]})$,  the file $W_{\theta}$ can be decoded  by \eqref{model:decod const}, i.e., $H(W_{\theta}|{A}_{[1:N]\backslash\mathcal{N}}^{[\theta]},{W}_{\mathcal{K}},{Z}_{\mathcal{N}},{Q}_{1:N}^{[\theta]})=0$;
$(d)$ is due to $H(W_\theta|{W}_{\mathcal{K}},{Z}_{\mathcal{N}},{Q}_{1:N}^{[\theta]})=H(W_\theta| {Z}_{\mathcal{N}})$ for $\theta\notin\mathcal{K}$ by the independence of the files \eqref{model:file inden} and the fact that queries are independent of the files \eqref{model:query inden};
$(e)$ and $(f)$ follow by simply changing the summation indices;
$(g)$ follows from $\sum\limits_{\mathcal{K}\subseteq[1:K]\backslash\{\theta\},|\mathcal{K}|=k} H(W_{\theta}|{Z}_{\mathcal{N}})=\binom{K-1}{k}H(W_{\theta}|{Z}_{\mathcal{N}})$;
$(h)$ follows from the definition of $\lambda_{n}$ in \eqref{Defi:lambda}.

Notice that when $n\in[0:N-1]$ and $k=K-1$, all the files $W_{1:K}$ are presented in  the conditions of each  entropy function in $\widetilde{T}(n,k)$. Since the answers $A_{[1:N]\backslash\mathcal{N}}^{[\theta]}$ is a function of $Q_{[1:N]\backslash\mathcal{N}}^{[\theta]}$ and ${W}_{1:K}$ by \eqref{model:answers}, we have $\widetilde{T}(n,K-1)=0$. Thus, by \eqref{appen:lower bound1}, for $n\in[0:N-1]$ and $k=K-1$,
\begin{IEEEeqnarray}{c}
T(n,K-1)\geq\frac{1}{N-n}\cdot\lambda_n,
\end{IEEEeqnarray}
This proves \eqref{Lemma:rescu} for $n\in[0:N-1]$ and $k=K-1$. We proceed to prove the other cases by deriving a lower bound on $\widetilde{T}(n,k)$.


Let $\mathcal{P}_N$ be the set consisting of all possible permutations of $[1:N]$, and  $\boldsymbol{\sigma}\triangleq(\sigma_{1},\sigma_{2},\ldots,\sigma_{N})\in \mathcal{P}_N$ denote a permutation operation of $[1:N]$.
For any $n\in[0:N-1]$ and $k\in[0:K-2]$, we further lower bound $\widetilde{T}(n,k)$ as follows.
\begin{IEEEeqnarray}{rCl}
&& \widetilde{T}(n,k) \notag\\
&=& \frac{1}{NK\tbinom{K-1}{k}\tbinom{N-1}{n}}\sum\limits_{\substack{\mathcal{K}\subseteq[1:K]\\|\mathcal{K}|=k+1}}\sum\limits_{\theta\in\mathcal{K}} \sum\limits_{\substack{\mathcal{N}\subseteq[1:N]\\|\mathcal{N}|=n}}
H({A}_{[1:N]\backslash\mathcal{N}}^{[\theta]}|{W}_{\mathcal{K}},{Z}_{\mathcal{N}},{Q}_{1:N}^{[\theta]}) \notag \\
&\overset{(a)}{=}& \frac{1}{NK\tbinom{K-1}{k}\tbinom{N-1}{n}}\sum\limits_{\substack{\mathcal{K}\subseteq[1:K]\\|\mathcal{K}|=k+1}} \sum\limits_{\theta\in\mathcal{K}} \frac{1}{n!(N-n)!}\sum\limits_{\boldsymbol{\sigma}\in \mathcal{P}_{N}} H({A}_{{\sigma}_{[n+1:N]}}^{[\theta]}|{W}_{\mathcal{K}},{Z}_{{\sigma}_{[1:n]}},{Q}_{1:N}^{[\theta]}) \notag \\
&\overset{(b)}{=}&\frac{1}{N!K\tbinom{K-1}{k}(N-n)} \sum\limits_{\substack{\mathcal{K}\subseteq[1:K]\\|\mathcal{K}|=k+1}} \sum\limits_{\theta\in\mathcal{K}} \sum\limits_{\boldsymbol{\sigma}\in \mathcal{P}_{N}} \sum\limits_{n'=n}^{N-1}
H(A_{{\sigma}_{n'+1}}^{[\theta]}|{W}_{\mathcal{K}},{Z}_{{\sigma}_{[1:n]}},{A}_{{\sigma}_{[n+1:n']}}^{[\theta]},{Q}_{1:N}^{[\theta]}) \notag \\
&\overset{(c)}{\geq}&\frac{1}{N!K\tbinom{K-1}{k}(N-n)} \sum\limits_{\substack{\mathcal{K}\subseteq[1:K]\\|\mathcal{K}|=k+1}} \sum\limits_{\theta\in\mathcal{K}} \sum\limits_{\boldsymbol{\sigma}\in \mathcal{P}_{N}} \sum\limits_{n'=n}^{N-1}
H(A_{{\sigma}_{n'+1}}^{[\theta]}|{W}_{\mathcal{K}},{Z}_{{\sigma}_{[1:n]}},{Z}_{{\sigma}_{[n+1:n']}},{A}_{{\sigma}_{[n+1:n']}}^{[\theta]},{Q}_{1:N}^{[\theta]}) \label{pro:Lemma2} \\
&\overset{(d)}{=}&\frac{1}{N!K\tbinom{K-1}{k}(N-n)} \sum\limits_{\substack{\mathcal{K}\subseteq[1:K]\\|\mathcal{K}|=k+1}} \sum\limits_{\theta\in\mathcal{K}} \sum\limits_{\boldsymbol{\sigma}\in \mathcal{P}_{N}} \sum\limits_{n'=n}^{N-1}
H(A_{{\sigma}_{n'+1}}^{[\theta]}|{W}_{\mathcal{K}},{Z}_{{\sigma}_{[1:n]}},{Z}_{{\sigma}_{[n+1:n']}},{Q}_{1:N}^{[\theta]}) \notag \\
&\overset{(e)}{=}&\frac{1}{N!K\tbinom{K-1}{k}(N-n)} \sum\limits_{\substack{\mathcal{K}\subseteq[1:K]\\|\mathcal{K}|=k+1}} \sum\limits_{\theta\in\mathcal{K}} \sum\limits_{n'=n}^{N-1}\sum\limits_{\boldsymbol{\sigma}\in \mathcal{P}_{N}}
H(A_{{\sigma}_{n'+1}}^{[\theta]}|{W}_{\mathcal{K}},{Z}_{{\sigma}_{[1:n']}},Q_{{\sigma}_{n'+1}}^{[\theta]}) \notag \\
&\overset{(f)}{=}&\frac{1}{N!K\tbinom{K-1}{k}(N-n)} \sum\limits_{\substack{\mathcal{K}\subseteq[1:K]\\|\mathcal{K}|=k+1}} \sum\limits_{\theta\in\mathcal{K}} \sum\limits_{n'=n}^{N-1} n'!(N-n'-1)! \sum\limits_{\substack{\mathcal{N}\subseteq[1:N]\\|\mathcal{N}|=n'}}\sum\limits_{i\in[1:N]\backslash\mathcal{N}}
H(A_{i}^{[\theta]}|{W}_{\mathcal{K}},{Z}_{\mathcal{N}},Q_{i}^{[\theta]}) \notag \\
&\overset{(g)}{=}&\frac{1}{N!K\tbinom{K-1}{k}(N-n)} \sum\limits_{\substack{\mathcal{K}\subseteq[1:K]\\|\mathcal{K}|=k+1}} \sum\limits_{\theta\in\mathcal{K}} \sum\limits_{n'=n}^{N-1} n'!(N-n'-1)! \sum\limits_{\substack{\mathcal{N}\subseteq[1:N]\\|\mathcal{N}|=n'}}\sum\limits_{i\in[1:N]\backslash\mathcal{N}}
\sum\limits_{\theta'\in[1:K]\backslash\mathcal{K}} \frac{H(A_{i}^{[\theta']}|{W}_{\mathcal{K}},{Z}_{\mathcal{N}},Q_{i}^{[\theta']})}{K-k-1} \notag \\
&\overset{(h)}{=}&\frac{1}{N!K\tbinom{K-1}{k}(N-n)} \sum\limits_{\substack{\mathcal{K}\subseteq[1:K]\\|\mathcal{K}|=k+1}}  \sum\limits_{n'=n}^{N-1} n'!(N-n'-1)! \sum\limits_{\substack{\mathcal{N}\subseteq[1:N]\\|\mathcal{N}|=n'}}\sum\limits_{i\in[1:N]\backslash\mathcal{N}}
\sum\limits_{\theta'\in[1:K]\backslash\mathcal{K}} \frac{k+1}{K-k-1}H(A_{i}^{[\theta']}|{W}_{\mathcal{K}},{Z}_{\mathcal{N}},Q_{i}^{[\theta']}) \notag \\
&=&\frac{1}{N-n} \sum\limits_{n'=n}^{N-1} \frac{1}{NK\tbinom{K-1}{k+1}\tbinom{N-1}{n'}} \sum\limits_{\substack{\mathcal{K}\subseteq[1:K]\\|\mathcal{K}|=k+1}} \sum\limits_{\substack{\mathcal{N}\subseteq[1:N]\\|\mathcal{N}|=n'}}\sum\limits_{\theta'\in[1:K]\backslash\mathcal{K}}  \sum\limits_{i\in[1:N]\backslash\mathcal{N}}  H(A_{i}^{[\theta']}|{W}_{\mathcal{K}},{Z}_{\mathcal{N}},Q_{i}^{[\theta']}) \notag \\
&=&\frac{1}{N-n} \sum\limits_{n'=n}^{N-1}T(n',k+1), \label{appen:lower bound2}
\end{IEEEeqnarray}
where $(a)$ holds because for each $\mathcal{N}\subseteq[1:N]$ with $|\mathcal{N}|=n$, there are exactly $n!(N-n)!$ permutation operations $\boldsymbol{\sigma}\in \mathcal{P}_{N}$ satisfying $H({A}_{{\sigma}_{[n+1:N]}}^{[\theta]}|{W}_{\mathcal{K}},{Z}_{{\sigma}_{[1:n]}},{Q}_{1:N}^{[\theta]})=H({A}_{[1:N]\backslash\mathcal{N}}^{[\theta]}|{W}_{\mathcal{K}},{Z}_{\mathcal{N}},{Q}_{1:N}^{[\theta]})$, i.e., such $n!(N-n)!$ permutation operations $\boldsymbol{\sigma}$ are restricted over $\mathcal{N}$ and $[1:N]\backslash\mathcal{N}$, hence,
\begin{IEEEeqnarray*}{c}
\sum\limits_{\boldsymbol{\sigma}\in \mathcal{P}_{N}} H({A}_{{\sigma}_{[n+1:N]}}^{[\theta]}|{W}_{\mathcal{K}},{Z}_{{\sigma}_{[1:n]}},{Q}_{1:N}^{[\theta]})
=n!(N-n)!\cdot\sum\limits_{\substack{\mathcal{N}\subseteq[1:N]\\|\mathcal{N}|=n}}H({A}_{[1:N]\backslash\mathcal{N}}^{[\theta]}|{W}_{\mathcal{K}},{Z}_{\mathcal{N}},{Q}_{1:N}^{[\theta]});
\end{IEEEeqnarray*}
$(b)$ follows from the chain rule of entropy; $(c)$ holds because conditioning reduces entropy; $(d)$ holds because the answers ${A}_{{\sigma}_{[n+1:n']}}^{[\theta]}$ is a function of the stored contents ${Z}_{{\sigma}_{[n+1:n']}}$ and the queries ${Q}_{{\sigma}_{[n+1:n']}}^{[\theta]}$ by \eqref{model:answers}; $(e)$ follows by \eqref{Appen:query}; $(f)$ follows from similar arguments to $(a)$; $(g)$ follows from \eqref{Appen:priva proof};  $(h)$ is due to $\sum_{\theta\in\mathcal{K}}H(A_{i}^{[\theta']}|{W}_{\mathcal{K}},{Z}_{\mathcal{N}},Q_{i}^{[\theta']})=(k+1)H(A_{i}^{[\theta']}|{W}_{\mathcal{K}},{Z}_{\mathcal{N}},Q_{i}^{[\theta']})$.

We prove \eqref{Lemma:rescu} by replacing \eqref{appen:lower bound1} with \eqref{appen:lower bound2}.

Remarkably, to establish the equality in \eqref{Lemma:rescu}, the inequalities in \eqref{pro:Lemma1} and \eqref{pro:Lemma2} have to hold with equalities.
\begin{itemize}
  \item The equality in \eqref{pro:Lemma1}  indicates that  for any $\mathcal{K}\subseteq[1:K]$ of size $k$, $\mathcal{N}\subseteq[1:N]$ of size $n$, and $\theta\in[1:K]\backslash\mathcal{K}$,
\begin{IEEEeqnarray}{c}\label{equal:1}
\sum\limits_{i\in[1:N]\backslash\mathcal{N}} H(A_{i}^{[\theta]}|{W}_{\mathcal{K}},{Z}_{\mathcal{N}},{Q}_{1:N}^{[\theta]})
= H({A}_{[1:N]\backslash\mathcal{N}}^{[\theta]}|{W}_{\mathcal{K}},{Z}_{\mathcal{N}},{Q}_{1:N}^{[\theta]}). \notag
\end{IEEEeqnarray}
That is,
\begin{IEEEeqnarray}{rCl}
0&=&\sum\limits_{i\in[1:N]\backslash\mathcal{N}} H(A_{i}^{[\theta]}|{W}_{\mathcal{K}},{Z}_{\mathcal{N}},{Q}_{1:N}^{[\theta]})- H({A}_{[1:N]\backslash\mathcal{N}}^{[\theta]}|{W}_{\mathcal{K}},{Z}_{\mathcal{N}},{Q}_{1:N}^{[\theta]}) \notag\\
&=&\sum\limits_{\widetilde{{Q}}_{1:N}^{[\theta]}}\Pr({Q}_{1:N}^{[\theta]}=\widetilde{{Q}}_{1:N}^{[\theta]})
\left[  \sum\limits_{i\in[1:N]\backslash\mathcal{N}} H(A_{i}^{[\theta]}|{W}_{\mathcal{K}},{Z}_{\mathcal{N}},{Q}_{1:N}^{[\theta]}=\widetilde{{Q}}_{1:N}^{[\theta]})- H({A}_{[1:N]\backslash\mathcal{N}}^{[\theta]}|{W}_{\mathcal{K}},{Z}_{\mathcal{N}},{Q}_{1:N}^{[\theta]}=\widetilde{{Q}}_{1:N}^{[\theta]}) \right]. \IEEEeqnarraynumspace \label{equal:11}
\end{IEEEeqnarray}
Whereas, for each realization of queries  $\widetilde{{Q}}_{1:N}^{[\theta]}$ with positive probability,
\begin{IEEEeqnarray}{c}\notag
\sum\limits_{i\in[1:N]\backslash\mathcal{N}} H(A_{i}^{[\theta]}|{W}_{\mathcal{K}},{Z}_{\mathcal{N}},{Q}_{1:N}^{[\theta]}=\widetilde{{Q}}_{1:N}^{[\theta]})\geq H({A}_{[1:N]\backslash\mathcal{N}}^{[\theta]}|{W}_{\mathcal{K}},{Z}_{\mathcal{N}},{Q}_{1:N}^{[\theta]}=\widetilde{{Q}}_{1:N}^{[\theta]}).
\end{IEEEeqnarray}
That is, the terms in square bracket of \eqref{equal:11} are nonnegative. Accordingly, \eqref{C11} holds for all realizations of queries  $\widetilde{{Q}}_{1:N}^{[\theta]}$ with positive probability.
\item Similarly, the equality in \eqref{pro:Lemma2} indicates that
for any $\mathcal{K}\subseteq[1:K], \theta\in\mathcal{K},\mathcal{N}_1\subseteq[1:N],\mathcal{N}_2\subseteq[1:N]\backslash\mathcal{N}_1$ and $i\in[1:N]\backslash(\mathcal{N}_1\cup\mathcal{N}_2)$ such that $|\mathcal{K}|=k+1$, $|\mathcal{N}_1|=n$, $|\mathcal{N}_2|=n'-n$ with $n'\in[n+1:N-1]$,
\begin{IEEEeqnarray}{rCl}
0&=&H(A_{i}^{[\theta]}|{W}_{\mathcal{K}},{Z}_{\mathcal{N}_1},{A}_{\mathcal{N}_2}^{[\theta]},{Q}_{1:N}^{[\theta]})
-H(A_{i}^{[\theta]}|{W}_{\mathcal{K}},{Z}_{\mathcal{N}_1},{Z}_{\mathcal{N}_2},{A}_{\mathcal{N}_2}^{[\theta]},{Q}_{1:N}^{[\theta]}) \notag \\
&=&I(A_{i}^{[\theta]};{Z}_{\mathcal{N}_2}|{W}_{\mathcal{K}},{Z}_{\mathcal{N}_1},{A}_{\mathcal{N}_2}^{[\theta]},{Q}_{1:N}^{[\theta]})\notag\\
&=&\sum\limits_{\widetilde{{Q}}_{1:N}^{[\theta]}}\Pr({Q}_{1:N}^{[\theta]}=\widetilde{{Q}}_{1:N}^{[\theta]})
I(A_{i}^{[\theta]};{Z}_{\mathcal{N}_2}|{W}_{\mathcal{K}},{Z}_{\mathcal{N}_1},{A}_{\mathcal{N}_2}^{[\theta]},{Q}_{1:N}^{[\theta]}=\widetilde{{Q}}_{1:N}^{[\theta]}). \label{equal:22}
\end{IEEEeqnarray}
Notice that the mutual information terms in \eqref{equal:22} are nonnegative. Consequently, they have to be zero for all realizations of queries $\widetilde{{Q}}_{1:N}^{[\theta]}$ with positive probability, i.e., \eqref{C22} holds.
\end{itemize}

The proof of this lemma is completed.
\end{IEEEproof}

Define   coefficients $\widetilde{\alpha}_{\ell}$ and a function $\widetilde{D}(\ell)$  as follows:
\begin{IEEEeqnarray}{rCl}
\widetilde{\alpha}_{\ell}&\triangleq&\frac{1}{\tbinom{N}{\ell}}\sum\limits_{\substack{\mathcal{S}\subseteq[1:N]\\|\mathcal{S}|=\ell}}\alpha_{\mathcal{S}}, \quad\forall\, \ell\in[0:N],\label{defi:xt}\\
\widetilde{D}(\ell)&\triangleq&1+\frac{1}{\ell}+\ldots+\frac{1}{\ell^{K-1}},\quad\forall\, \ell\in[1:N+1]\label{def:dt}
\end{IEEEeqnarray}
with the boundary conditions
\begin{IEEEeqnarray}{c}\notag
\widetilde{\alpha}_{N+1}\triangleq0,\quad\widetilde{D}(0)\triangleq NK.
\end{IEEEeqnarray}
We next apply $\widetilde{\alpha}_\ell$ to write the constraint of file size in \eqref{const:file size} as
\begin{IEEEeqnarray}{c}
\sum\limits_{\ell=1}^{N}\binom{N}{\ell}\widetilde{\alpha}_\ell=1.\label{proof:file const}
\end{IEEEeqnarray}

\begin{Lemma}\label{cost:down lower bound}
The download cost $D$ for any SC-PIR scheme has the following lower bound:
\begin{IEEEeqnarray}{c}
D\geq L\cdot\sum\limits_{\ell=1}^{N}\binom{N}{\ell}\widetilde{D}(\ell)\widetilde{\alpha}_{\ell}.\label{lemma:download}
\end{IEEEeqnarray}
Moreover, given any $\mathcal{S}\subseteq[1:N]$ and $\theta,\theta'\in[1:K]$ such that $|\mathcal{S}|=M\in[2:N]$ and $\theta\neq\theta'$, if the equality in \eqref{lemma:download} holds, then
for every realization of queries  $\widetilde{{Q}}_{1:N}^{[\theta]}$ with positive probability, any SC-PIR scheme must satisfy
\begin{itemize}
  \item[P3$'$.] The answers at servers in $\mathcal{S}$ are independent of each other in the conditioning of ${W}_{[1:K]\backslash\{\theta\}}$ and ${Z}_{[1:N]\backslash\mathcal{S}}$, i.e.,
\begin{IEEEeqnarray}{c}\notag
\sum\limits_{i\in\mathcal{S}} H(A_{i}^{[\theta]}|{W}_{[1:K]\backslash\{\theta\}},{Z}_{[1:N]\backslash\mathcal{S}},{Q}_{1:N}^{[\theta]}=\widetilde{{Q}}_{1:N}^{[\theta]})
= H({A}_{\mathcal{S}}^{[\theta]}|{W}_{[1:K]\backslash\{\theta\}},{Z}_{[1:N]\backslash\mathcal{S}},{Q}_{1:N}^{[\theta]}=\widetilde{{Q}}_{1:N}^{[\theta]}).
\end{IEEEeqnarray}
\item[P4$'$.] The answers at servers in $\mathcal{S}$ are independent of each other in the conditioning of ${W}_{\theta'}$ and ${Z}_{[1:N]\backslash\mathcal{S}}$, i.e.,
\begin{IEEEeqnarray}{c}\notag
\sum\limits_{i\in\mathcal{S}} H(A_{i}^{[\theta]}|{W}_{\theta'},{Z}_{[1:N]\backslash\mathcal{S}},{Q}_{1:N}^{[\theta]}=\widetilde{{Q}}_{1:N}^{[\theta]})
= H({A}_{\mathcal{S}}^{[\theta]}|{W}_{\theta'},{Z}_{[1:N]\backslash\mathcal{S}},{Q}_{1:N}^{[\theta]}=\widetilde{{Q}}_{1:N}^{[\theta]}).
\end{IEEEeqnarray}
\item[P5$'$.] The answer at server $i$ is independent of the contents stored at server $j$ in the conditioning of ${W}_{\theta},{W}_{\theta'},{Z}_{[1:N]\backslash\mathcal{S}}$ and ${A}_{j}^{[\theta]}$ for any $i,j\in\mathcal{S}$, i.e.,
\begin{IEEEeqnarray}{c}\notag
I(A_{i}^{[\theta]};{Z}_{j}|{W}_{\theta},{W}_{\theta'},{Z}_{[1:N]\backslash\mathcal{S}},{A}_{j}^{[\theta]},{Q}_{1:N}^{[\theta]}=\widetilde{{Q}}_{1:N}^{[\theta]})=0,\quad\forall\, i,j\in\mathcal{S}.
\end{IEEEeqnarray}
\end{itemize}
\end{Lemma}
\begin{IEEEproof}
We have the following two boundary conditions on $T(n,k)$ in \eqref{defi:Tnk} and $\lambda_{n}$ in \eqref{Defi:lambda}:
\begin{IEEEeqnarray}{c}\label{condi:intial1}
T(0,0)=\frac{1}{NK}\sum\limits_{\theta\in[1:K]}\sum\limits_{i\in[1:N]}H(A_{i}^{[\theta]}|Q_{i}^{[\theta]})
\overset{(a)}{=}\frac{1}{N}\sum\limits_{i\in[1:N]}H(A_{i}^{[\theta]}|Q_{i}^{[\theta]}),
\end{IEEEeqnarray}
where $(a)$ follows from \eqref{Appen:priva proof} by setting $\mathcal{N}=\emptyset$ and $\mathcal{K}=\emptyset$, and
\begin{IEEEeqnarray}{c}\label{condi:intial3}
\lambda_{0}=\frac{1}{K}\sum\limits_{\theta\in[1:K]}H(W_{\theta})\overset{(a)}{=}L,
\end{IEEEeqnarray}
where $(a)$ is due to \eqref{infor indenpe}.
Therefore,
\begin{IEEEeqnarray}{rCl}
D&\overset{(a)}{=}&\sum_{i=1}^{N}H(A_{i}^{[\theta]})\notag\\
&\geq&\sum\limits_{i=1}^{N}H(A_{i}^{[\theta]}|Q_{i}^{[\theta]}) \notag \\
&\overset{(b)}{=}&N\cdot T(0,0) \notag \\
&\overset{(c)}{\geq}&\lambda_0+\sum\limits_{n_1=0}^{N-1}T(n_1,1) \notag\\
&\overset{(d)}{\geq}& \lambda_{0}+\sum\limits_{n_{1}=0}^{N-1}\frac{\lambda_{n_1}}{N-n_1}+\sum\limits_{n_{1}=0}^{N-1}\sum\limits_{n_{2}=n_{1}}^{N-1}\frac{\lambda_{n_{2}}}{(N-n_{1})(N-n_{2})}
+\sum\limits_{n_{1}=0}^{N-1}\sum\limits_{n_{2}=n_{1}}^{N-1}\sum\limits_{n_{3}=n_{2}}^{N-1}\frac{\lambda_{n_{3}}}{(N-n_{1})(N-n_{2})(N-n_{3})} \notag \\
&&+\ldots+\sum\limits_{n_{1}=0}^{N-1}\sum\limits_{n_2=n_1}^{N-1}\ldots\sum\limits_{n_{K-1}=n_{K-2}}^{N-1}\frac{\lambda_{n_{K-1}}+\sum_{n_{K}=n_{K-1}}^{N-1}T(n_{K},K)}{(N-n_{1})\times(N-n_{2})\times\ldots\times(N-n_{K-1})} \notag \\
&\overset{(e)}{=}& L+\sum\limits_{n_{1}=0}^{N-1}\frac{\lambda_{n_1}}{N-n_1}+\sum\limits_{n_{1}=0}^{N-1}\sum\limits_{n_{2}=n_{1}}^{N-1}\frac{\lambda_{n_{2}}}{(N-n_{1})(N-n_{2})}
+\sum\limits_{n_{1}=0}^{N-1}\sum\limits_{n_{2}=n_{1}}^{N-1}\sum\limits_{n_{3}=n_{2}}^{N-1}\frac{\lambda_{n_{3}}}{(N-n_{1})(N-n_{2})(N-n_{3})} \notag \\
&&+\ldots+\sum\limits_{n_{1}=0}^{N-1}\sum\limits_{n_2=n_1}^{N-1}\ldots\sum\limits_{n_{K-1}=n_{K-2}}^{N-1}\frac{\lambda_{n_{K-1}}}{(N-n_{1})\times(N-n_{2})\times\ldots\times(N-n_{K-1})} \notag \\
&\overset{(f)}{=}& L+\sum\limits_{n_{1}=1}^{N}\frac{\lambda_{N-n_1}}{n_1}+\sum\limits_{n_{1}=1}^{N}\sum\limits_{n_{2}=1}^{n_{1}}\frac{\lambda_{N-n_{2}}}{n_{1}n_{2}}
+\sum\limits_{n_{1}=1}^{N}\sum\limits_{n_{2}=1}^{n_{1}}\sum\limits_{n_{3}=1}^{n_{2}}\frac{\lambda_{N-n_{3}}}{n_{1}n_{2}n_{3}} +\ldots+\sum\limits_{n_{1}=1}^{N}\ldots\sum\limits_{n_{K-1}=1}^{n_{K-2}}\frac{\lambda_{N-n_{K-1}}}{n_{1}\times\ldots\times n_{K-1}} \notag \\
&\overset{(g)}{=}& L+\sum\limits_{n_{1}=1}^{N}\frac{\lambda_{N-n_1}}{n_1}+\sum\limits_{n_{1}=1}^{N}\sum\limits_{n_{2}=n_{1}}^{N}\frac{\lambda_{N-n_{1}}}{n_{1}n_{2}}
+\sum\limits_{n_{1}=1}^{N}\sum\limits_{n_{2}=n_{1}}^{N}\sum\limits_{n_{3}=n_{2}}^{N}\frac{\lambda_{N-n_{1}}}{n_{1}n_{2}n_{3}} +\ldots+\sum\limits_{n_{1}=1}^{N}\ldots\sum\limits_{n_{K-1}=n_{K-2}}^{N}\frac{\lambda_{N-n_{1}}}{n_{1}\times\ldots\times n_{K-1}} \notag \\
&=&L+\sum\limits_{n_1=1}^{N}\Bigg( \underbrace{\frac{1}{n_1}+\sum\limits_{n_{2}=n_{1}}^{N}\frac{1}{n_{1}n_{2}}+\sum\limits_{n_{2}=n_{1}}^{N}\sum\limits_{n_{3}=n_{2}}^{N}\frac{1}{n_1n_2n_3}
+\ldots+\sum\limits_{n_{2}=n_{1}}^{N}\ldots\sum\limits_{n_{K-1}=n_{K-2}}^{N}\frac{1}{n_1\times\ldots\times n_{K-1}}}_{\triangleq S(n_1,K)} \Bigg)\cdot\lambda_{N-n_1} \notag \\
&\overset{(h)}{=}& L+\sum\limits_{n_1=1}^{N}\sum\limits_{\ell=1}^{n_1}\binom{n_1}{\ell}S(n_1,K)\widetilde{\alpha}_{\ell}L \notag \\
&=&L\cdot \Bigg( 1+\sum\limits_{\ell=1}^{N}\widetilde{\alpha}_{\ell}\underbrace{\sum\limits_{n_1=\ell}^{N}\binom{n_1}{\ell}S(n_1,K)}_{\triangleq\alpha(\ell,K)} \Bigg) \notag\\
&\overset{(i)}{=}&L\cdot \Bigg( 1+\sum\limits_{\ell=1}^{N}\binom{N}{\ell}\left( \widetilde{D}(\ell)-1 \right)\widetilde{\alpha}_{t} \Bigg) \notag\\
&\overset{(j)}{=}&L\cdot  \sum\limits_{\ell=1}^{N}\binom{N}{\ell}\widetilde{D}(\ell)\widetilde{\alpha}_{\ell}, \notag
\end{IEEEeqnarray}
where $(a)$ follows from \eqref{def:rate}; $(b)$ follows from  \eqref{condi:intial1};
$(c)$ and $(d)$ follow by applying \eqref{Lemma:rescu} $K$ times recursively;
$(e)$ follows from \eqref{condi:intial3} and \eqref{condi:intial2};
$(f)$ and $(g)$ follow by changing of the summation indices simply;
$(h)$ holds because $\lambda_n=\sum_{\ell=1}^{N-n}\binom{N-n}{\ell}\widetilde{\alpha}_{\ell} L$ by the result in \cite[Eq. (54)]{Attia SC-PIR};
$(i)$ follows from the definition of $\widetilde{D}(\ell)$ in \eqref{def:dt} and $\alpha(\ell,K)=\tbinom{N}{\ell}(\widetilde{D}(\ell)-1 )$ by the result in \cite[Eq. (60)]{Attia SC-PIR};
$(j)$ follows from \eqref{proof:file const}.

In order to establish the equality in \eqref{lemma:download}, it is easy to see that, in the steps of applying \eqref{Lemma:rescu}, all the inequalities must hold with equalities. This implies that the equality in \eqref{Lemma:rescu} holds for all $n\in[0:N-1]$ and $k\in[1:K-1]$. Thus, by Lemma \ref{recr:lemma}, \eqref{C11} and \eqref{C22} hold for any $n\in[0:N-1]$ and $k\in[1:K-1]$. 


Accordingly, for any $\theta,\theta'\in[1:K]$  and $\mathcal{S}\subseteq[1:N]$ such that  $\theta\neq \theta'$ and $|\mathcal{S}|=N-n=M$, we have
\begin{itemize}
\item  Let  $\mathcal{K}=[1:K]\backslash\{\theta\}$ and $\mathcal{N}=[1:N]\backslash\mathcal{S}$, then $\mathcal{K}$ is of size $k=K-1$ and $\mathcal{N}$ is of size $n\in[0:N-1]$ (i.e., $M\in[1:N]$). Thus, by \eqref{C11}, 
\begin{IEEEeqnarray}{c}\notag
\sum\limits_{i\in\mathcal{S}} H(A_{i}^{[\theta]}|{W}_{[1:K]\backslash\{\theta\}},{Z}_{[1:N]\backslash\mathcal{S}},{Q}_{1:N}^{[\theta]}=\widetilde{{Q}}_{1:N}^{[\theta]})
=H({A}_{\mathcal{S}}^{[\theta]}|{W}_{[1:K]\backslash\{\theta\}},{Z}_{[1:N]\backslash\mathcal{S}},{Q}_{1:N}^{[\theta]}=\widetilde{{Q}}_{1:N}^{[\theta]}).
\end{IEEEeqnarray}
\item  Let $\mathcal{K}=\{\theta'\}$ and $\mathcal{N}=[1:N]\backslash\mathcal{S}$,  then $\mathcal{K}$ is of size $k=1$ and $\mathcal{N}$ is of size $n\in[0:N-1]$. Thus, by \eqref{C11} again, 
\begin{IEEEeqnarray}{c}\notag
\sum\limits_{i\in\mathcal{S}} H(A_{i}^{[\theta]}|{W}_{\theta'},{Z}_{[1:N]\backslash\mathcal{S}},{Q}_{1:N}^{[\theta]}=\widetilde{{Q}}_{1:N}^{[\theta]})
=H({A}_{\mathcal{S}}^{[\theta]}|{W}_{\theta'},{Z}_{[1:N]\backslash\mathcal{S}},{Q}_{1:N}^{[\theta]}=\widetilde{{Q}}_{1:N}^{[\theta]}).
\end{IEEEeqnarray}
\item For $n\in[0:N-2]$ (i.e., $M\in[2:N]$), let $k=1$ and $n'=n+1$, then  $\mathcal{K}=\{\theta,\theta'\}$ is of size $k+1$, $\mathcal{N}_1=[1:N]\backslash\mathcal{S}$ is of size $n$, $\mathcal{N}_2=\{j\}$ is of size $n'-n=1$ for $j\in\mathcal{S}$. Thus given any $i\in\mathcal{S}\backslash\{j\}$, by \eqref{C22}, 
\begin{IEEEeqnarray}{c}\notag
I(A_{i}^{[\theta]};{Z}_{j}|{W}_{\theta},{W}_{\theta'},{Z}_{[1:N]\backslash\mathcal{S}},{A}_{j}^{[\theta]},{Q}_{1:N}^{[\theta]}=\widetilde{{Q}}_{1:N}^{[\theta]})=0.
\end{IEEEeqnarray}
\end{itemize}
To conclude, P3$'$-P5$'$ must hold if the equality in \eqref{lemma:download} holds. 
\end{IEEEproof}

\subsection{Proof of Lemma \ref{necessary:P1-2}}

For fixed $j\in[0:N]$,
the total storage  of the $N$ servers \eqref{uncode:storage constraint} is constrained as
\begin{IEEEeqnarray}{rCl}\label{proof:stro const}
\mu N&\geq&\sum\limits_{n\in[1:N]}\sum\limits_{\substack{\mathcal{S}\subseteq[1:N]\\n\in\mathcal{S}}}\alpha_{\mathcal{S}} \label{inequality:muN} \\
&{=}& \sum\limits_{\ell=1}^{N} \ell\binom{N}{\ell}  \widetilde{\alpha}_\ell \notag \\
&=& j+\sum\limits_{\ell\in[1:N]\backslash\{j,j+1\}}\binom{N}{\ell}(\ell-j)\widetilde{\alpha}_\ell+\binom{N}{j+1}\widetilde{\alpha}_{j+1}, \label{proof:storage}
\end{IEEEeqnarray}
 where the last two equalities follow from \eqref{defi:xt} and \eqref{proof:file const}, respectively.
Hence,
\begin{IEEEeqnarray}{rCl}
\frac{D}{L}&\overset{(a)}{\geq}& \sum\limits_{\ell=1}^{N}\binom{N}{\ell}\widetilde{D}(\ell)\widetilde{\alpha}_\ell  \label{inequality:a}\\
&\overset{(b)}{=}&\widetilde{D}(j)+\sum\limits_{\ell\in[1:N]\backslash\{j,j+1\}}\binom{N}{\ell}\left(\widetilde{D}(\ell)-\widetilde{D}(j)\right)\widetilde{\alpha}_\ell+\binom{N}{j+1}\left(\widetilde{D}(j+1)-\widetilde{D}(j)\right)\widetilde{\alpha}_{j+1}\notag\\
&\overset{(c)}{\geq}& \widetilde{D}(j)+\sum\limits_{\ell\in[1:N]\backslash\{j,j+1\}}\binom{N}{\ell}\left( \widetilde{D}(\ell)-\widetilde{D}(j)\right)\widetilde{\alpha}_\ell
+\left( \widetilde{D}(j+1)-\widetilde{D}(j) \right)\Big( \mu N-j-\sum\limits_{\ell\in[1:N]\backslash\{j,j+1\}}\binom{N}{\ell}(\ell-j)\widetilde{\alpha}_\ell\Big)  \label{inequality:c} \IEEEeqnarraynumspace\\
&=&(\mu N-j)\widetilde{D}(j+1)-(\mu N-j-1)\widetilde{D}(j)+\sum\limits_{\ell\in[1:N]\backslash\{j,j+1\}}\binom{N}{\ell}\Big(\widetilde{D}(\ell)+(\ell-j-1)\widetilde{D}(j)-(\ell-j)\widetilde{D}(j+1)\Big)\widetilde{\alpha}_\ell\notag\\
&\overset{(d)}{\geq}& (\mu N-j)\widetilde{D}(j+1)-(\mu N-j-1)\widetilde{D}(j) \label{proof:xt} \\
&=& (M-j)\widetilde{D}(j+1)-(M-j-1)\widetilde{D}(j), \label{proof:result}
\end{IEEEeqnarray}
where $(a)$ follows by \eqref{lemma:download};
$(b)$ is due to \eqref{proof:file const};
$(c)$ follows from \eqref{proof:storage} and the fact that $\widetilde{D}(j+1)-\widetilde{D}(j)$ is negative for all $j\in[0:N]$;
$(d)$ is because $\widetilde{\alpha}_{\ell}\geq 0$ and  $\widetilde{D}(\ell)+(\ell-j-1)\widetilde{D}(j)-(\ell-j)\widetilde{D}(j+1)\geq 0$  for all $\ell\in[1:N]\backslash\{j,j+1\}$ by \cite[Lemma 5]{Attia SC-PIR}.

Therefore, by \eqref{def:rate} and \eqref{proof:result}, we have
\begin{IEEEeqnarray}{c}\label{proof:rate}
R\leq \left( (M-j)\widetilde{D}(j+1)-(M-j-1)\widetilde{D}(j) \right)^{-1},\quad\forall\,j\in[0:N].
\end{IEEEeqnarray}

In particular, for $j=M$ and $j=M-1$, \eqref{proof:rate} results in
\begin{IEEEeqnarray}{c}
R\leq \left(\widetilde{D}(M) \right)^{-1}=\left( 1+\frac{1}{M}+\ldots+\frac{1}{M^{K-1}} \right)^{-1},\label{Bound capaci}
\end{IEEEeqnarray}
which achieves the capacity of the SC-PIR system (see \eqref{optimal download}).

 Thus, for any capacity-achieving  SC-PIR scheme, the inequality in \eqref{Bound capaci} must  hold with equality, which implies that the inequalities in \eqref{inequality:a}, \eqref{inequality:c} and \eqref{proof:xt}  hold with equality for both $j=M$ and $j=M-1$. Therefore,
 \begin{itemize}
   \item  It is easy to prove $\widetilde{D}(\ell)+(\ell-j-1)\widetilde{D}(j)-(\ell-j)\widetilde{D}(j+1)>0$ for all $\ell\in[1:N]\backslash\{j,j+1\}$ by \cite[Lemma 5]{Attia SC-PIR}. Moreover, we also have $\widetilde{\alpha}_\ell\geq 0$.
   Thus, the equality in \eqref{proof:xt} for $j=M$ and $j=M-1$ indicates that $\widetilde{\alpha}_\ell=0$ holds for all $\ell\in([1:N]\backslash\{M,M+1\})\cup([1:N]\backslash\{M-1,M\})=[1:N]\backslash\{M\}$. Accordingly, by \eqref{defi:xt}, we have $\alpha_{\mathcal{S}}=0$ for all $\mathcal{S}\subseteq[1:N]$ with $|\mathcal{S}|\neq M$.
   \item Due to $\widetilde{D}(j+1)-\widetilde{D}(j)<0$ for all $j\in[0:N]$, thus the equality in \eqref{inequality:c} indicates that the equality in \eqref{inequality:muN} also holds, i.e., the equalities in \eqref{uncode:storage constraint} hold for all $n\in[1:N]$.
 \end{itemize}

As a result, P1-P2 must be satisfied in the storage design phase of any capacity-achieving SC-PIR scheme.
 In addition,  the equality in \eqref{inequality:a} indicates $D=L\cdot\sum_{\ell=1}^{N}\tbinom{N}{\ell}\widetilde{D}(\ell)\widetilde{\alpha}_\ell$, thus we obtain the following corollary by Lemma \ref{cost:down lower bound}.
\begin{Corollary}\label{corollary:lemma1}
Any capacity-achieving SC-PIR scheme must satisfy P3$'$-P5$'$. 
\end{Corollary}

\subsection{Proof of Lemma \ref{lemma:necess}}
By Corollary \ref{corollary:lemma1}, P3$'$-P5$'$ hold for any capacity-achieving SC-PIR scheme. Next, we prove Lemma \ref{lemma:necess} by specializing P3$'$-P5$'$ to linear SC-PIR schemes.

According to Lemma \ref{necessary:P1-2}, in the capacity-achieving SC-PIR scheme, all the servers  only store the packets placing in $M$ different servers, thus the stored contents \eqref{storage place} at server $n$ are simplified to
\begin{IEEEeqnarray}{C}\label{capacity:storage}
Z_{n}=\mathop\cup\limits_{k\in[1:K]}\mathop\cup\limits_{\substack{\mathcal{S}\subseteq[1:N]\\|\mathcal{S}|=M,n\in\mathcal{S}}}{W}_{k,\mathcal{S}},\quad\forall\, n\in[1:N].
\end{IEEEeqnarray}

For any realization of queries $\widetilde{{Q}}_{1:N}^{[\theta]}$ with positive probability, the answer $A_{n}^{[\theta]}$ is a deterministic function of the corresponding query realization $\widetilde{Q}_{n}^{[\theta]}$ and the stored contents $Z_{n}$ by \eqref{model:answers}. Therefore, by the stored contents in \eqref{capacity:storage}, $A_{n}^{[\theta]}$ is merely the function of $W_{\theta,\mathcal{S}}$ and $\widetilde{Q}_{n}^{[\theta]}$ conditioned on the files ${W}_{[1:K]\backslash\{\theta\}}$ and ${Z}_{[1:N]\backslash\mathcal{S}}$. For linear SC-PIR scheme, it is exactly $\widetilde{\textbf{LC}}_{n}^{[\theta]}(W_{\theta,\mathcal{S}})$ conditioned on ${W}_{[1:K]\backslash\{\theta\}}$ and ${Z}_{[1:N]\backslash\mathcal{S}}$. Therefore,
\begin{IEEEeqnarray}{rCl}
&&\sum\limits_{n\in\mathcal{S}} H\Big(A_{n}^{[\theta]}\,\big|\,{W}_{[1:K]\backslash\{\theta\}},{Z}_{[1:N]\backslash\mathcal{S}},{Q}_{1:N}^{[\theta]}=\widetilde{{Q}}_{1:N}^{[\theta]}\Big) \notag\\
&=& \sum\limits_{n\in\mathcal{S}} H\Big(\widetilde{\textbf{LC}}_{n}^{[\theta]}(W_{\theta,\mathcal{S}})\,\big|\,{W}_{[1:K]\backslash\{\theta\}},{Z}_{[1:N]\backslash\mathcal{S}},{Q}_{1:N}^{[\theta]}=\widetilde{{Q}}_{1:N}^{[\theta]}\Big) \notag\\
&\overset{(a)}{=}& \sum\limits_{n\in\mathcal{S}} H\big(\widetilde{\textbf{LC}}_{n}^{[\theta]}(W_{\theta,\mathcal{S}})\big), \label{lowerbound:p3}
\end{IEEEeqnarray}
where $(a)$ follows from independence of all the packets \eqref{packets:independent}, and the fact that queries are independent of files by \eqref{model:query inden}.

Similarly,
\begin{IEEEeqnarray}{rCl}
&& H\Big({A}_{\mathcal{S}}^{[\theta]}\,\big|\,{W}_{[1:K]\backslash\{\theta\}},{Z}_{[1:N]\backslash\mathcal{S}},{Q}_{1:N}^{[\theta]}=\widetilde{{Q}}_{1:N}^{[\theta]}\Big) \notag\\
&=& H\Big(\widetilde{\textbf{LC}}_{n}^{[\theta]}(W_{\theta,\mathcal{S}}):n\in\mathcal{S}\,\big|\,{W}_{[1:K]\backslash\{\theta\}},{Z}_{[1:N]\backslash\mathcal{S}},{Q}_{1:N}^{[\theta]}=\widetilde{{Q}}_{1:N}^{[\theta]}\Big) \notag\\
&=& H\Big(\widetilde{\textbf{LC}}_{n}^{[\theta]}(W_{\theta,\mathcal{S}}):n\in\mathcal{S}\Big).\label{lowerbound:p3a}
\end{IEEEeqnarray}
Substituting the two sides in P3$'$ for \eqref{lowerbound:p3} and \eqref{lowerbound:p3a}, we obtain
\begin{IEEEeqnarray}{c}\notag
\sum\limits_{n\in\mathcal{S}} H\Big(\widetilde{\textbf{LC}}_{n}^{[\theta]}(W_{\theta,\mathcal{S}})\Big)=
H\Big(\widetilde{\textbf{LC}}_{n}^{[\theta]}(W_{\theta,\mathcal{S}}):n\in\mathcal{S}\Big).
\end{IEEEeqnarray}
Hence the random variables $\big\{\widetilde{\textbf{LC}}_{n}^{[\theta]}(W_{\theta,\mathcal{S}}):n\in\mathcal{S}\big\}$ are independent of each other, i.e.,  P3 holds.

By the similar argument to \eqref{lowerbound:p3}, P4$'$ can be rewritten as
\begin{IEEEeqnarray}{c}\notag
\sum\limits_{n\in\mathcal{S}} H\Big(\widetilde{\mathbf{LC}}_{n}^{[\theta]}(W_{[1:K]\backslash\{\theta'\},\mathcal{S}})\Big)=
H\Big(\widetilde{\mathbf{LC}}_{n}^{[\theta]}(W_{[1:K]\backslash\{\theta'\},\mathcal{S}}):n\in\mathcal{S}\Big).
\end{IEEEeqnarray}
Therefore,
\begin{IEEEeqnarray}{c}\notag
\widetilde{\mathbf{LC}}_{n}^{[\theta]}\big(W_{[1:K]\backslash\{\theta'\},\mathcal{S}}\big),\quad\forall\, n\in\mathcal{S}
\end{IEEEeqnarray}
are also independent of each other, i.e., P4 holds.

Let $n,n'\in\mathcal{S}$ and $\theta'\neq \theta$. By P5$'$,
\begin{IEEEeqnarray}{rCl}
0 &=& I\Big(A_{n}^{[\theta]};{Z}_{n'}\,\big|\,{W}_{\theta},{W}_{\theta'},{Z}_{[1:N]\backslash\mathcal{S}},{A}_{n'}^{[\theta]},{Q}_{1:N}^{[\theta]}=\widetilde{{Q}}_{1:N}^{[\theta]}\Big) \notag\\
  &\overset{(a)}{=}& I\Big(\widetilde{\textbf{LC}}_{n}^{[\theta]}(W_{[1:K]\backslash\{\theta,\theta'\},\mathcal{S}});W_{[1:K]\backslash\{\theta,\theta'\},\mathcal{S}}\,\big|\,\widetilde{\mathbf{LC}}_{n'}^{[\theta]}(W_{[1:K]\backslash\{\theta,\theta'\},\mathcal{S}}) \Big) \notag\\
&=& H\Big(\widetilde{\textbf{LC}}_{n}^{[\theta]}(W_{[1:K]\backslash\{\theta,\theta'\},\mathcal{S}})\,\big|\,\widetilde{\textbf{LC}}_{n'}^{[\theta]}(W_{[1:K]\backslash\{\theta,\theta'\},\mathcal{S}}) \Big) \notag\\ &&\quad\quad-H\Big(\widetilde{\mathbf{LC}}_{n}^{[\theta]}(W_{[1:K]\backslash\{\theta,\theta'\},\mathcal{S}}\}\,\big|\,W_{[1:K]\backslash\{\theta,\theta'\},\mathcal{S}},\widetilde{\mathbf{LC}}_{n'}^{[\theta]}(W_{[1:K]\backslash\{\theta,\theta'\},\mathcal{S}}) \Big) \notag\\
&\overset{(b)}{=}&H\Big(\widetilde{\mathbf{LC}}_{n}^{[\theta]}(W_{[1:K]\backslash\{\theta,\theta'\},\mathcal{S}})\,\big|\,\widetilde{\mathbf{LC}}_{n'}^{[\theta]}(W_{[1:K]\backslash\{\theta,\theta'\},\mathcal{S}})\Big), \notag
\end{IEEEeqnarray}
where $(a)$ follows from the similar argument to \eqref{lowerbound:p3} again; $(b)$ is due to the fact that $\widetilde{\mathbf{LC}}_{n}^{[\theta]}(W_{[1:K]\backslash\{\theta,\theta'\},\mathcal{S}})$ is a function of packets in  $W_{[1:K]\backslash\{\theta,\theta'\},\mathcal{S}}$ by \eqref{add:notation1} and \eqref{add:notation2}. Thus,
\begin{IEEEeqnarray}{c}\notag
\widetilde{\mathbf{LC}}_{n}^{[\theta]}\big(W_{[1:K]\backslash\{\theta,\theta'\},\mathcal{S}}\big),\quad\forall\, n\in\mathcal{S}
\end{IEEEeqnarray}
are deterministic of each other, i.e., P5 holds.

As a result, P3-P5 are necessary conditions of any capacity-achieving linear SC-PIR scheme.



\end{document}